\documentclass[preprint,12pt]{elsarticle}

\usepackage{amsmath}
\usepackage{amssymb}
\usepackage{graphicx}
\usepackage{caption}
\usepackage{array}
\usepackage{epstopdf}
\usepackage{hyperref}
\usepackage{booktabs}
\usepackage{geometry}
\usepackage{subcaption}
\usepackage{threeparttable}

\usepackage{color,soul}
\usepackage{xcolor}
\sethlcolor{yellow}

\colorlet{lightcyan}{cyan!30}    
\colorlet{lightgreen}{lime!30}   
\colorlet{lightorange}{orange!35} 

\begin{document}

\begin{frontmatter}

\title{PyPSA-Libya: An Open-Source Energy System Model for National Decarbonization and Hydrogen Export Pathways}

\author{Mohamed Abuella\corref{cor1}}
\ead{mhdabuella@gmail.com}
\cortext[cor1]{Any feedback is welcome via email.}

\affiliation{organization={Independent Researcher},
             city={Newcastle upon Tyne},
             country={United Kingdom}}


\begin{abstract}
Libya possesses significant renewable energy potential that remains largely untapped. This paper presents PyPSA-Libya, a national-scale model built on the PyPSA-Earth framework, designed to explore optimal transition pathways for the Libyan power sector. We present two primary studies: (1) an electricity-only capacity expansion analysis comparing a zero-emission constraint against an unconstrained baseline at 36-cluster resolution, and (2) a sector-coupled hydrogen export study evaluating the levelized cost of exporting 70 TWh/year of hydrogen to Europe under three technological scenarios and a $100\text{~\texteuro/tonne}$ $\text{CO}_2$ carbon tax regime. Results indicate that while solar PV dominates the domestic capacity mix, integration of hydrogen production significantly alters grid storage requirements. Furthermore, under carbon pricing, steam methane reforming with carbon capture (SMR-CC) frequently outcompetes pure green electrolysis, establishing a lower-cost compliance pathway for cross-border export. The model, data, and configuration files are released under an open-source license to support transparent energy planning in Libya.
\end{abstract}


\end{frontmatter}

\section{Introduction}
\label{sec:introduction}

The global imperative to mitigate climate change has accelerated the transition from carbon-intensive fossil fuels to renewable energy sources, in line with the objectives of the Paris Agreement \cite{almaktar2021prospects, alanazi2025renewable}. Petro-state economies face a distinct dual challenge during this transition: they must decarbonize their domestic power sectors while simultaneously diversifying their economies away from oil and gas dependence \cite{almaktar2021prospects, eljrushi1990solar}. Libya represents a critical case study of this paradigm. The country's political economy relies almost entirely on conventional fossil fuels, with the oil sector contributing approximately 95\% of export earnings and 65\% of Gross Domestic Product (GDP) \cite{almaktar2021prospects}. Over the past decade, political instability and administrative division have severely affected Libya's energy sector. Armed conflicts and unilateral shutdowns of major oil fields (such as those in early 2020 causing financial losses exceeding \$9 billion) have repeatedly undermined the reliability of fuel supplies for electricity generation \cite{almaktar2021prospects}.

Concurrently, Libya's domestic power sector, monopolized by the state-owned General Electricity Company of Libya (GECOL), is facing a severe operational crisis \cite{almaktar2021prospects}. The power grid relies completely on fossil fuels, with natural gas driving approximately 63\% of generation and oil-fired plants accounting for 29\% \cite{almaktar2021prospects}. Rapid growth in energy demand, driven by rising temperatures and population growth, has led to a persistent peak-load electricity deficit of approximately 1,200~MW, resulting in frequent and widespread blackouts \cite{almaktar2021prospects}. Furthermore, Libya's environmental footprint is substantial; in 2018, its per capita CO2 emissions amounted to 8.73 tonnes, surpassing the Chinese national average of 7.95 tonnes \cite{almaktar2021prospects}. Although the Renewable Energy Authority of Libya (REAoL) set an ambitious target in 2010 to integrate a 10\% renewable share (equivalent to 2,219~MW) into the national mix by 2025, this initiative was completely derailed by the civil conflict that erupted in 2011 \cite{almaktar2021prospects}.

Despite these challenges, Libya is exceptionally well-positioned to pivot toward a sustainable energy economy. The country possesses some of the highest solar and wind resource potentials in the world. The average global horizontal irradiation (GHI) over the Libyan territory yields an average potential solar photovoltaic (PV) productivity of 1.9~MWh/kW/year (with Al Kufra in the southeast reaching an optimal annual productivity of 2,100~kWh/kWp), while onshore wind potentials reach up to 500~W/m$^2$ along its southernmost borders, coastal Darna, and the Western Mountain region \cite{almaktar2021prospects}. Recognizing these vast resources, pioneering research in the late 1980s first proposed a solar-hydrogen energy system for Libya, emphasizing that converting solar power to hydrogen would not only extend the lifespan of fluid fossil fuel reserves as raw chemical feedstocks but also position Libya as a leading exporter of clean hydrogen to Europe in lieu of oil and gas \cite{eljrushi1990solar}. 

To evaluate the modern technical and economic feasibility of this transition, we introduce \textsf{PyPSA-Libya}, a high-resolution, open-source energy system model of Libya's power and hydrogen network. Built as a country-specific extension of the global \textsf{PyPSA-Earth} framework \cite{abdelkhalek2025pypsa}, our model is divided into two primary, complementary studies designed to address both domestic and international planning questions:
\begin{enumerate}
    \item {Electricity-only capacity expansion (Part 1):} This study models Libya's power sector in isolation at a high spatial resolution of 36 clustered nodes. It compares a zero-CO2 emission constraint (Case 1) against an unconstrained fossil-heavy baseline (Case 2) under 2030 technology-cost assumptions, capturing transmission bottlenecks and identifying the optimal mix of solar, wind, and storage needed for complete domestic grid decarbonization.
    \item {Sector-coupled hydrogen export (Part 2):} This study models the spatial and economic implications of exporting 70~TWh/year of renewable or low-carbon hydrogen from Libya to European markets via maritime shipping \cite{abdelkhalek2025pypsa}. Formulated at a 12-cluster resolution to maintain computational tractability, we evaluate three technology scenarios representing strictly green electrolysis (Scenario A), electrolysis coupled with carbon-captured steam methane reforming (SMR CC) representing blue hydrogen (Scenario B), and an unconstrained scenario representing grey, blue, and green pathways (Scenario C).
\end{enumerate}

To ensure robustness, our model incorporates a \textsf{connectivity\_upgrade} workflow to resolve topological gaps and isolated buses inherent in OpenStreetMap (OSM) grid data for Libya. We report the Levelized Cost of Electricity (LCOE) and apply a paired-run differencing methodology to isolate the Net Levelized Cost of Hydrogen (Net LCOH) directly attributable to the export obligation. The model code, parameters, and results are released under an open-source license to establish a transparent, reproducible benchmark for national decarbonization and cross-border energy partnerships.

\section{Related work}
\label{sec:related_work}

The design of the \textsf{PyPSA-Libya} model builds upon three main pillars: (i) localized studies on Libyan renewable and hydrogen potentials, (ii) global hydrogen trade pathways and underground storage economics, and (iii) advancements in open-source spatial-temporal modeling and physical constraints.

\subsection{Renewable potential and hydrogen in Libya}
Early national energy assessments by Eljrushi and Veziroglu \cite{eljrushi1990solar} mathematically simulated the dynamic interaction between population growth and fossil fuel depletion, demonstrating that Libya would face severe energy supply deficits without a strategic transition toward solar-hydrogen production. Almaktar and Shaaban \cite{almaktar2021prospects} modernized these estimates through GIS and meteorological evaluations, assessing Libya's solar, wind, biomass, and geothermal resources as potential alternatives to conventional energy sources.

While macro-level studies established these resource potentials, localized techno-economic analyses remain rare. Rahil et al. \cite{rahil2018techno} evaluated off-peak forecourt electrolysis in Darnah under 2015 and 2030 cost horizons; however, their scope was restricted to isolated refueling stations without capturing nationwide transmission grid dynamics or spatial sector coupling. Our model bridges this gap by representing the full national transmission topology and co-optimizing domestic and export infrastructure.

\subsection{Global hydrogen trade and geological storage}
Cross-border hydrogen trade involves significant techno-economic and infrastructural uncertainties \cite{fakhreddine2025global}. In the Euro-Mediterranean context, TIAM-ECN modeling by Fattahi et al. \cite{fattahi2024opportunities} identified North Africa as the most cost-effective pipeline supplier of gaseous hydrogen to Europe, with liquefied shipping providing complementary supply diversification. Alanazi et al. \cite{alanazi2025renewable} confirmed that international trade could meet over 31\% of global decarbonized hydrogen demand by 2050, highlighting North Africa-to-Europe as a key trade corridor. Nevertheless, high conversion and transport losses dictate that hydrogen should be prioritized for hard-to-abate sectors, long-duration storage, and heavy industry rather than direct space heating or light mobility \cite{johnson2025realistic}.

Managing the volatility of renewable supply under rigid export commitments presents a major operational challenge \cite{terlouw2022large}. Achhammer et al. \cite{achhammer2026impact} addressed this by integrating GIS-mapped geological salt caverns into \textsf{PyPSA-Earth} for Tunisia, showing that underground hydrogen storage (UHS) reduces levelized export costs by ~3.7\%, dampens price volatility, and safeguards domestic energy security. \textsf{PyPSA-Libya} therefore assesses how the integration of maritime shipping logistics and underground storage limits can support resilient hydrogen exports.

\subsection{Open-source modeling and the PyPSA framework}
Proprietary energy planning models (e.g., TIMES, PLEXOS) often present commercial license barriers that limit transparency and collaboration in developing nations. To address this, the Python for Power System Analysis (PyPSA) library was introduced as a fully open-source framework for high-resolution capacity expansion planning. While \textsf{PyPSA-Eur} pioneered continental-scale transmission and storage optimization, country-specific forks such as \textsf{PyPSA-Spain} \cite{gallego2025pypsaspain} demonstrated the necessity of integrating detailed national datasets and nested border-price models to accurately reflect local grid conditions.

Expanding open-source modeling globally, Parzen et al. \cite{parzen2023pypsa} developed \textsf{PyPSA-Earth}, an automated global framework for extracting high-voltage network topologies from OpenStreetMap and mapping spatiotemporal weather potentials. Abdel-Khalek et al. \cite{abdelkhalek2025pypsa} extended this platform into a sector-coupled model incorporating multi-carrier energy vectors (electricity, hydrogen, gas, biomass, oil, and heat) across industry, transport, and residential sectors. Their application, which estimated Egypt's potential to supply 16\% of European hydrogen demand by 2030, provides a methodological reference for our sector-coupled export study. Similarly, Kwak et al. \cite{kwak2025pypsa} leveraged the \textsf{PyPSA-Earth} workflow to develop \textsf{PyPSA-Korea}, demonstrating how high-resolution open-source data can effectively resolve transmission bottlenecks and infrastructure sizing in isolated national grids.

Krog and Sperling \cite{krog2019comprehensive} emphasized the need for holistic strategic energy planning that integrates sustainability, multiple energy sectors, stakeholder and institutional coordination, and implementation across different planning levels to avoid sub-optimization. These governance and cross-sectoral principles highlight the necessity of moving beyond purely technical expansion planning.

However, default macro-energy optimization architectures still tend to evaluate technology placement strictly through a techno-economic lens, frequently omitting local physical and hydrological constraints such as inland water availability for electrolysis. To resolve this limitation, Walter et al.~\cite{walter2025integrating} integrated freshwater limits, reverse osmosis desalination, and water input constraints into energy system optimization models, showing that ignoring water availability leads to suboptimal electrolyser positioning and a ~10\% spatial capacity relocation across Europe. Integrating these physical and hydrological factors alongside cross-sectoral planning directly underpins the spatially resolved approach adopted in~\textsf{PyPSA-Libya}.


\section{Methodology}
\label{sec:methodology}

The~\textsf{PyPSA-Libya} model is developed using the open-source~\textsf{PyPSA-Earth} framework for large-scale energy system optimization~\cite{pypsa_earth_docs}. The model minimizes total annualized system costs, comprising capital expenditures (CAPEX) and variable operational costs (OPEX), across defined temporal and spatial horizons.

\begin{figure}[htbp]
    \centering
    \includegraphics[width=0.98\textwidth]{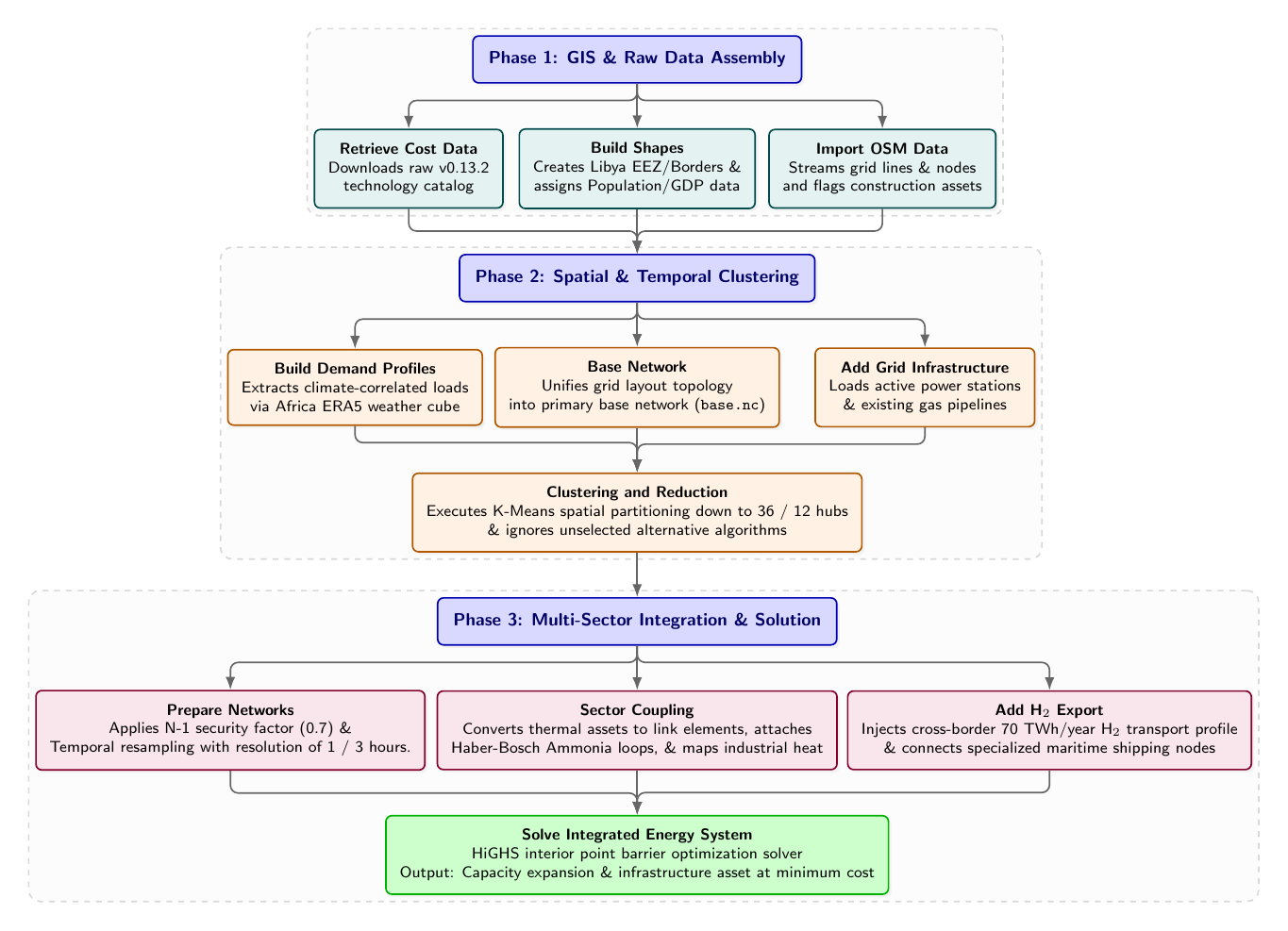}
    \caption{Computational workflow and data-processing pipeline of the PyPSA-Libya optimization framework, spanning from raw geographic data assembly to multi-sector matrix solution.}
    \label{fig:pypsa_libya_flowchart}
\end{figure}

The comprehensive computational and data-processing framework of the~\textsf{PyPSA-Libya} model is systematically structured into three core operational phases, as delineated in Figure~\ref{fig:pypsa_libya_flowchart}. 
The workflow advances from raw geographic GIS boundary extraction and brownfield asset assembly (Phase 1), through spatial K-means network partitioning and temporal discretization routines (Phase 2), and culminates in multi-carrier sector coupling, maritime export profile integration, and final large-scale Linear Programming matrix optimization via the HiGHS interior point solver (Phase 3).

\subsection{Input data and network topology}
The model utilizes OpenStreetMap (OSM) data for high-voltage transmission lines and substations. Preliminary diagnostic testing revealed significant data gaps in the Libyan OSM extract, specifically regarding isolated buses in the southern and desert regions (e.g., Fezzan and Al Kufra). To ensure a fully interconnected national grid backbone, synthetic line connections were added to guarantee that every node retains at least one physical connection to the transmission system.

Renewable resource potentials are derived from the ERA5 reanalysis weather dataset for the year 2013. The spatial resolution varies by study to balance detail with computational tractability:
\begin{enumerate}
    \item {Electricity-only study:} 36 clustered nodes, providing high-fidelity insight into internal transmission bottlenecks.
    \item {Sector-coupled study:} 12 clustered nodes, optimized to ensure the complex linear programming (LP) problem fits within 16~GB of memory.
\end{enumerate}

\subsection{Techno-economic assumptions}
All technology cost assumptions are calibrated to a 2030 planning horizon using the \textsf{PyPSA} technology-data repository. We assume a standard Weighted Average Cost of Capital (WACC) of 7.1\% (discount rate). The temporal resolution is resampled to 3-hour steps for the electricity study and 6-hour steps for the hydrogen study. This resolution is identified as the ``sweet spot'' for capturing solar diurnal ramps and battery cycling while preventing the numerical instability observed in 1-hour resolution runs for large-scale matrices.

\subsection{Scenario design}
\label{ssec:scenarios_design}
\begin{figure}[htbp]
    \centering
    \includegraphics[width=0.98\textwidth]{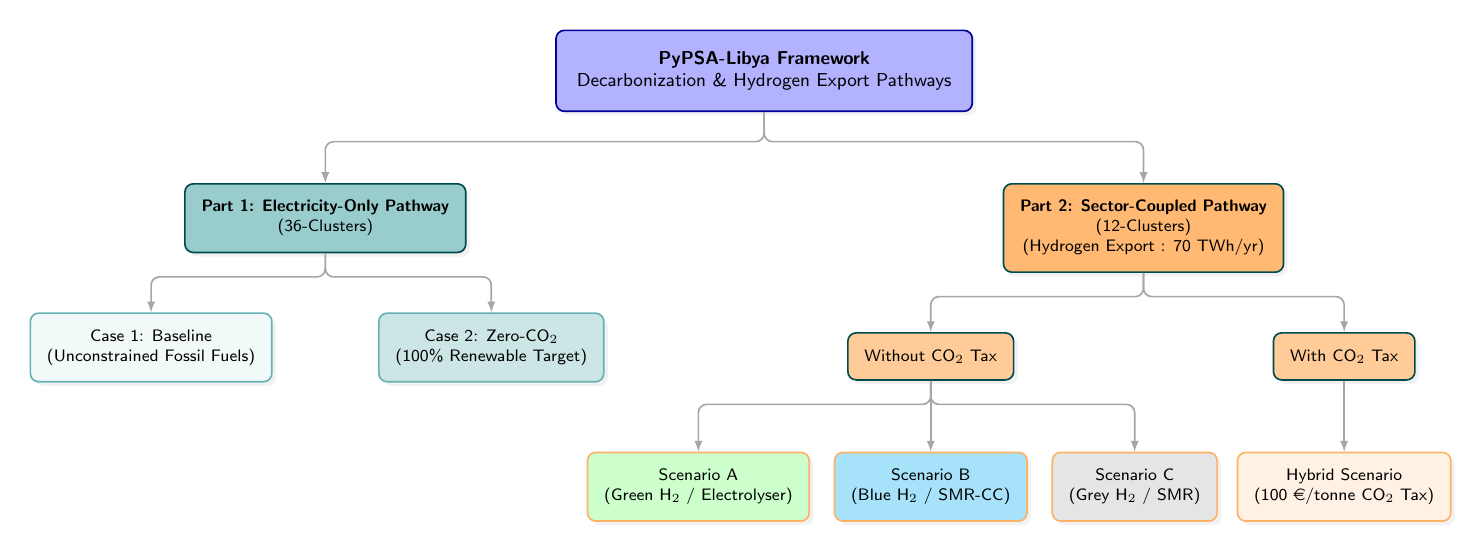}
    \caption{Schematic overview of the scenario design and modeling architecture for PyPSA-Libya.}
    \label{fig:pypsa_libya_methodology}
\end{figure}

The analytical workflow of this study is structured into a comprehensive two-part modeling framework to evaluate both domestic power sector decarbonization and large-scale cross-border hydrogen export pathways. As illustrated in Figure~\ref{fig:pypsa_libya_methodology}, the framework systematically bridges high-resolution national electricity grid optimization with multi-scenario sector-coupled hydrogen supply chains and policy regime optimizations.

\subsubsection{Part 1: Electricity-only decarbonization}
To establish a decarbonization baseline, we compare two cases within the electricity-only framework (no sector coupling):
\begin{itemize}
    \item {Case 1 (Baseline):} Imposes a non-binding cap ($1.0 \times 10^9$~tonnes), allowing the model to freely select the cost-minimal generation mix relying heavily on natural gas.
    \item {Case 2 (Zero-CO2):} Imposes a hard \textsf{co2\_limit} of 0.0, prohibiting all fossil dispatch (OCGT, CCGT, oil) and forcing a 100\% renewable system (Solar, Wind, Battery, and $\text{H}_2$ storage).
\end{itemize}

\subsubsection{Part 2: Sector-coupled hydrogen export}
Following the methodology of \cite{abdelkhalek2025pypsa}, we model the export of 70~TWh/year of hydrogen via maritime shipping. 

\begin{figure}[htbp]
    \centering
    \includegraphics[width=0.98\textwidth]{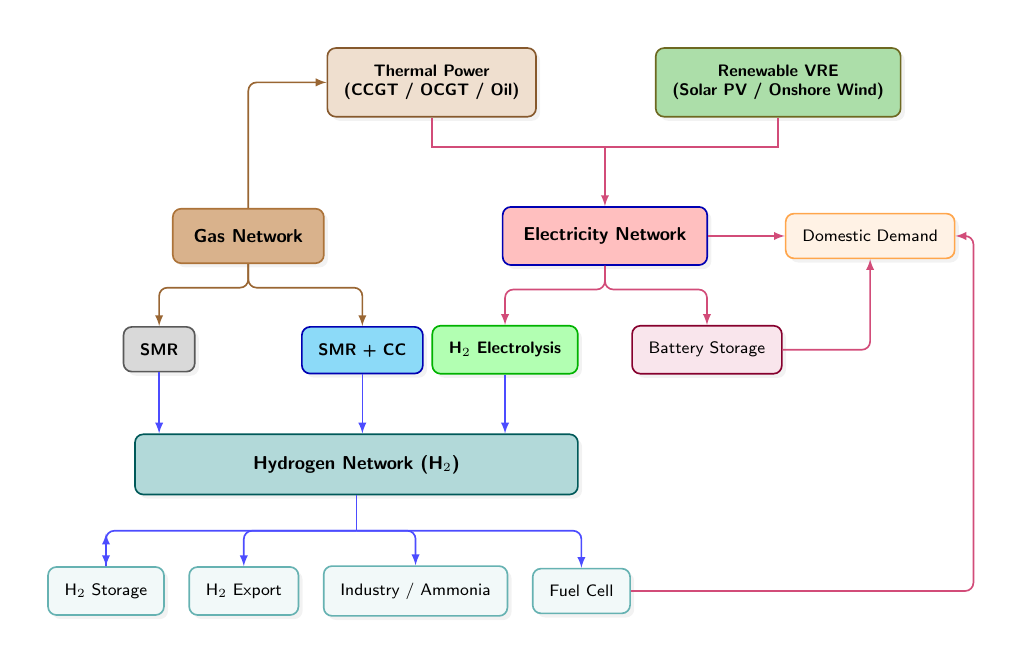}
    \caption{Multi-sector energy coupling and carrier flow architecture of the PyPSA-Libya model, illustrating the integrated network pathways across gas ($\text{CH}_4$), electricity (AC), and hydrogen ($\text{H}_2$) infrastructure.}
    \label{fig:sector_coupling_diag}
\end{figure}

The interconnected multi-sector coupling and carrier flow architecture implemented within the optimization framework is schematically illustrated in Figure~\ref{fig:sector_coupling_diag}. This model structure integrates the gas, electricity, and hydrogen networks into a single, cohesive topology, tracking simultaneous energy flows from primary extraction and generation units down to conversion interfaces, flexible storage systems, domestic demand centers, and international export channels.

The study is structured into two distinct policy and optimization phases to evaluate how carbon pricing alters technological choices:

\noindent{Phase 1: Baseline technology scenarios (No $\text{CO}_2$ tax)}
We first evaluate production economics under existing baseline market structures with zero penalty on carbon venting, isolating the performance of three distinct technological tracks:
\begin{itemize}
    \item {Scenario A (Green H$_2$):} Only water electrolysis is allowed, enforcing a strict annual renewable matching constraint and restricting electrolyser deployment to coastal nodes to utilize seawater desalination infrastructure.
    \item {Scenario B (Blue H$_2$):} Force-limits production to electrolysis and Steam Methane Reforming with Carbon Capture (SMR CC) under a rigid $\text{CO}_2$ cap ($1.0 \times 10^7$~tonnes).
    \item {Scenario C (Grey H$_2$):} An unconstrained baseline pathway allowing electrolysis, SMR CC, and unabated (grey) SMR under a non-binding emissions limit.
\end{itemize}

\noindent{Phase 2: Hybrid policy optimization (with $\text{CO}_2$ tax)}
We then introduce a uniform carbon tax applied directly to all net atmospheric emissions. Rather than running the previous scenarios in isolation, this phase removes all technological boundaries to form a single, integrated hybrid co-optimized scenario. In this unified run, the model is granted access to all three technology pathways simultaneously. The optimizer dynamically co-optimizes investments across green electrolysis, blue SMR CC, and grey SMR, automatically selecting the optimal hybrid technology mix that minimizes the total system cost under the new carbon tax regime.

\subsection{Solver configuration}
Numerical stability and computational scale are critical challenges in full-year, sector-coupled models. We utilize the HiGHS solver configured with the interior-point method (IPM). For complex runs, we disable the crossover phase and apply relaxed feasibility and optimality tolerances ($1 \times 10^{-3}$) to prevent numerical stall. The exact linear programming matrix dimensions, variable counts, and computational scaling metrics for Scenarios A, B, and C are detailed in~\ref{sec:appendix_problem_scale}.

\subsection{Data and cost assumptions}
We utilize ERA5 reanalysis weather data for the year 2013, resampled to 3-hour or 6-hour intervals. Technology costs are calibrated to the \textbf{2030 horizon}. A critical modification in \textsf{PyPSA-Libya} is the \textsf{connectivity\_upgrade} setting, which ensures that every node in the OSM-derived topology has at least one connection, mitigating the issue of isolated buses common in Libyan data.

\section{Results: electricity-only decarbonization}
\label{sec:results_electricity_only}

To evaluate the fundamental trade-offs of decarbonizing Libya's domestic power sector, this section presents the optimization results for the electricity-only framework (Part 1). We analyze the system's structural shift by contrasting Case 1 (Zero-$\text{CO}_2$), which enforces complete reliance on renewable energy and storage, against Case 2 (Baseline), where fossil dispatch remains unconstrained. The analysis evaluates spatial capacity expansion, transmission grid bottlenecks, system cost structures, and financial feasibility metrics to assess the technical and economic viability of a fully decarbonized national grid.

\subsection{Capacity of the energy mix}
\begin{figure}[htbp]
\centering
\begin{subfigure}[b]{0.49\textwidth}
    \centering
    \includegraphics[height=5.2cm]{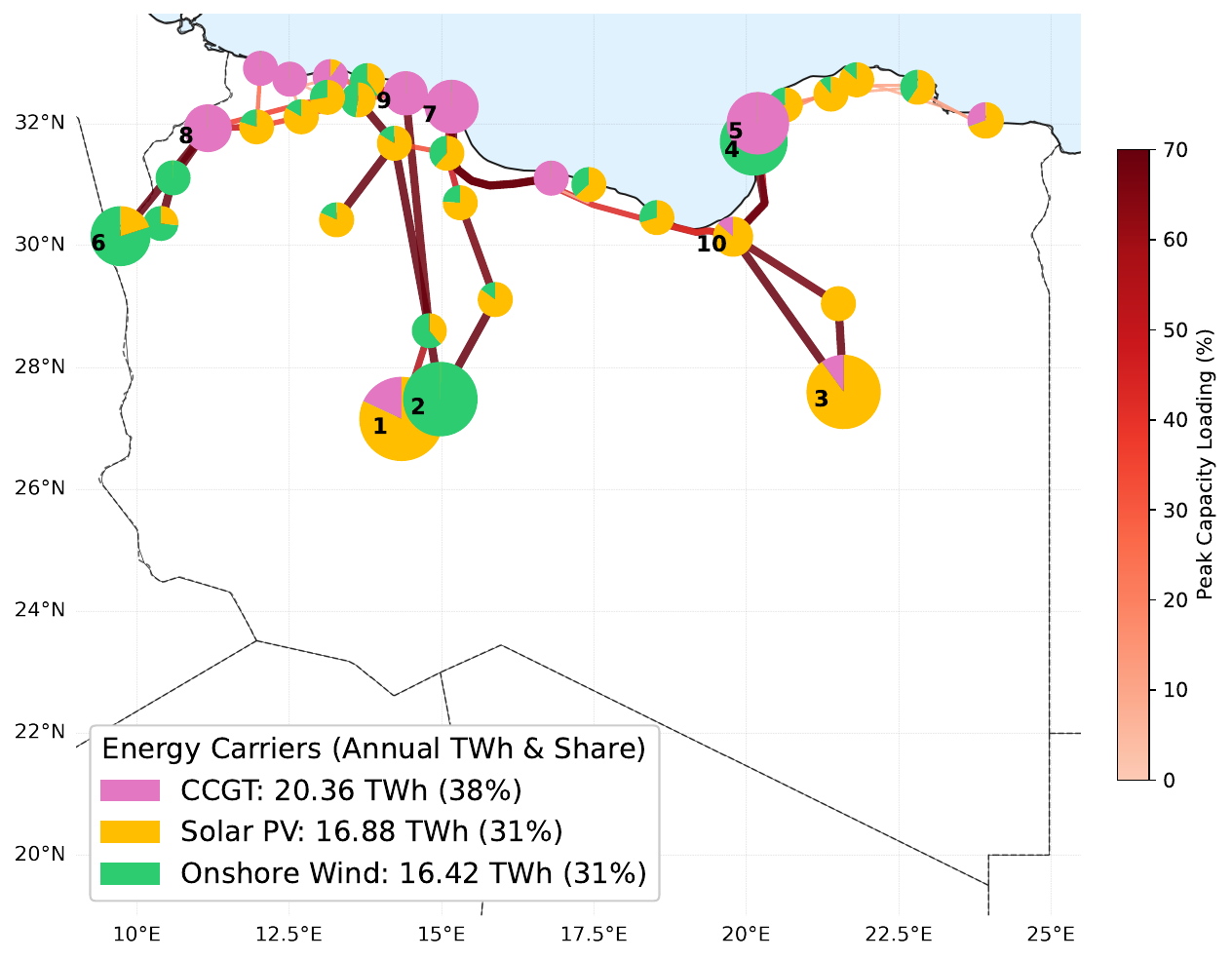}
    \caption{}
    \label{fig:No_LCO2_piecharts_map}
\end{subfigure}
\hfill
\begin{subfigure}[b]{0.49\textwidth}
    \centering
    \includegraphics[height=5.2cm]{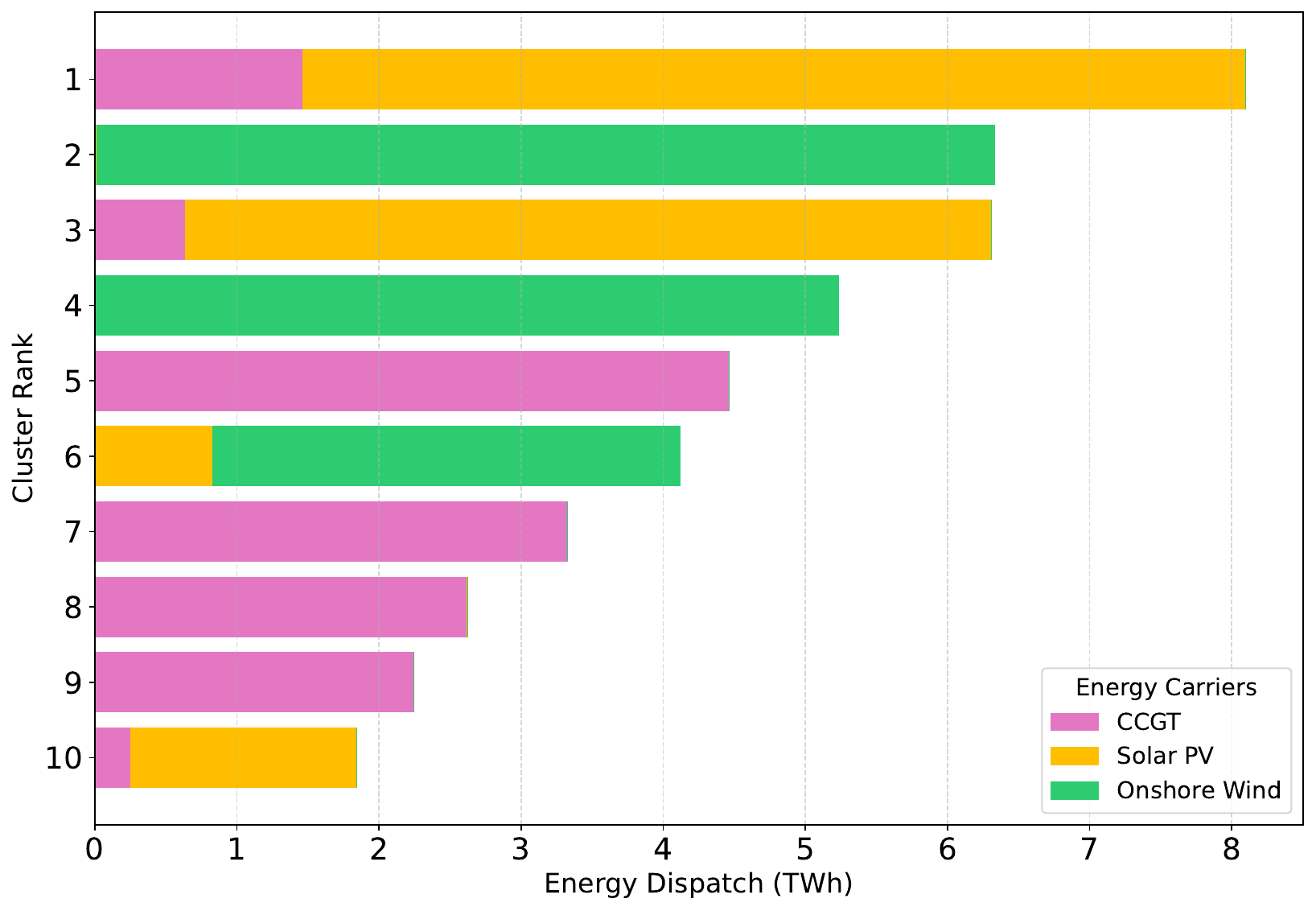}
    \caption{}
    \label{fig:No_LCO2_barchart}
\end{subfigure}
\caption{Optimal generation and transmission layout for the electricity-only model (ulimited CO$_2$, the baseline case): (a) Energy mix and annualized peak capacity loading of transmission lines across the clusters, (b) Distribution of the optimal generation and storage capacity mix for the top 10 clusters.}
\label{fig:No_LCO2_energy_mix}
\end{figure}

\begin{figure}[htbp]
\centering
\begin{subfigure}[b]{0.49\textwidth}
    \centering
    \includegraphics[height=5.2cm]{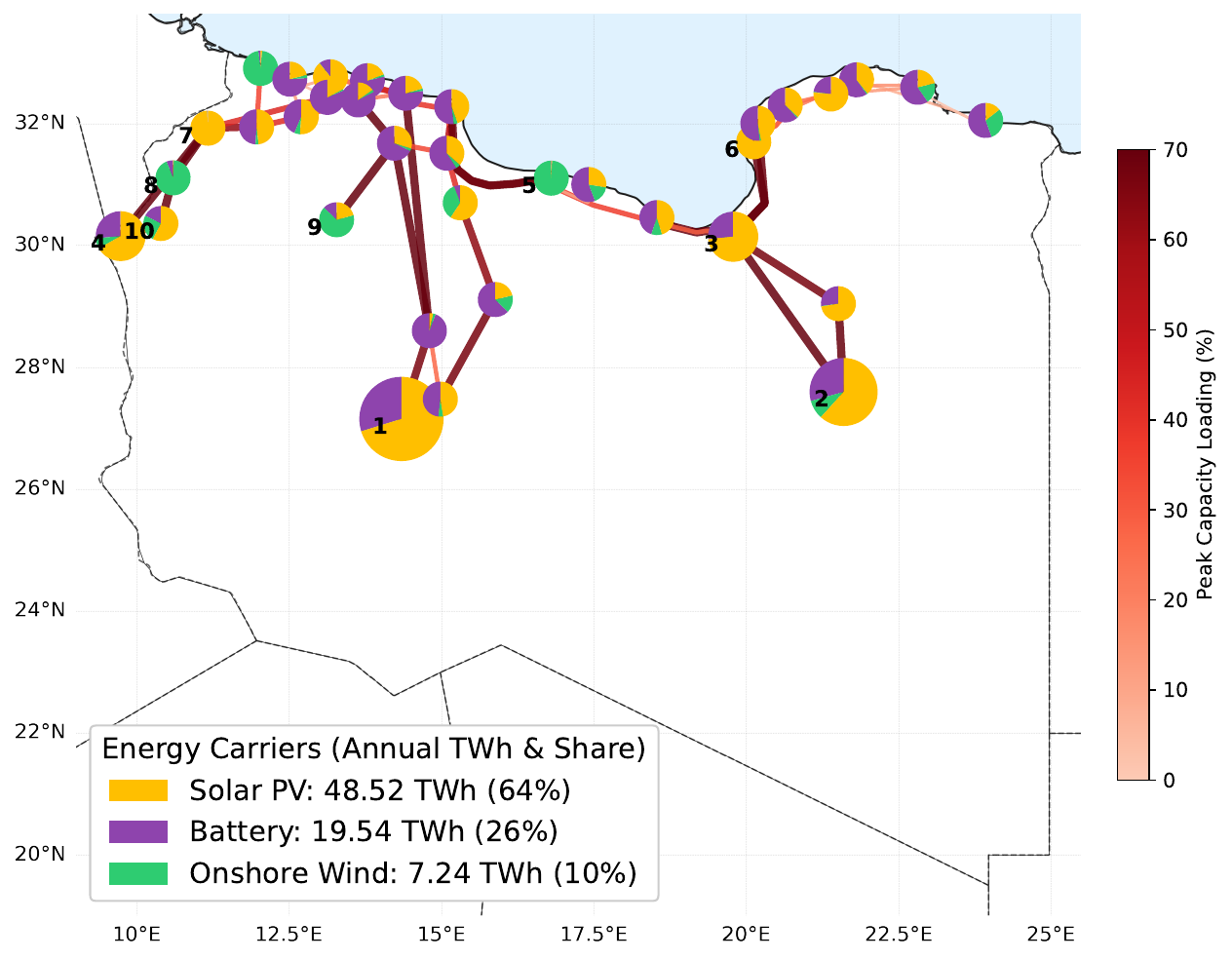}
    \caption{}
    \label{fig:LCO2_piecharts_map}
\end{subfigure}
\hfill
\begin{subfigure}[b]{0.49\textwidth}
    \centering
    \includegraphics[height=5.2cm]{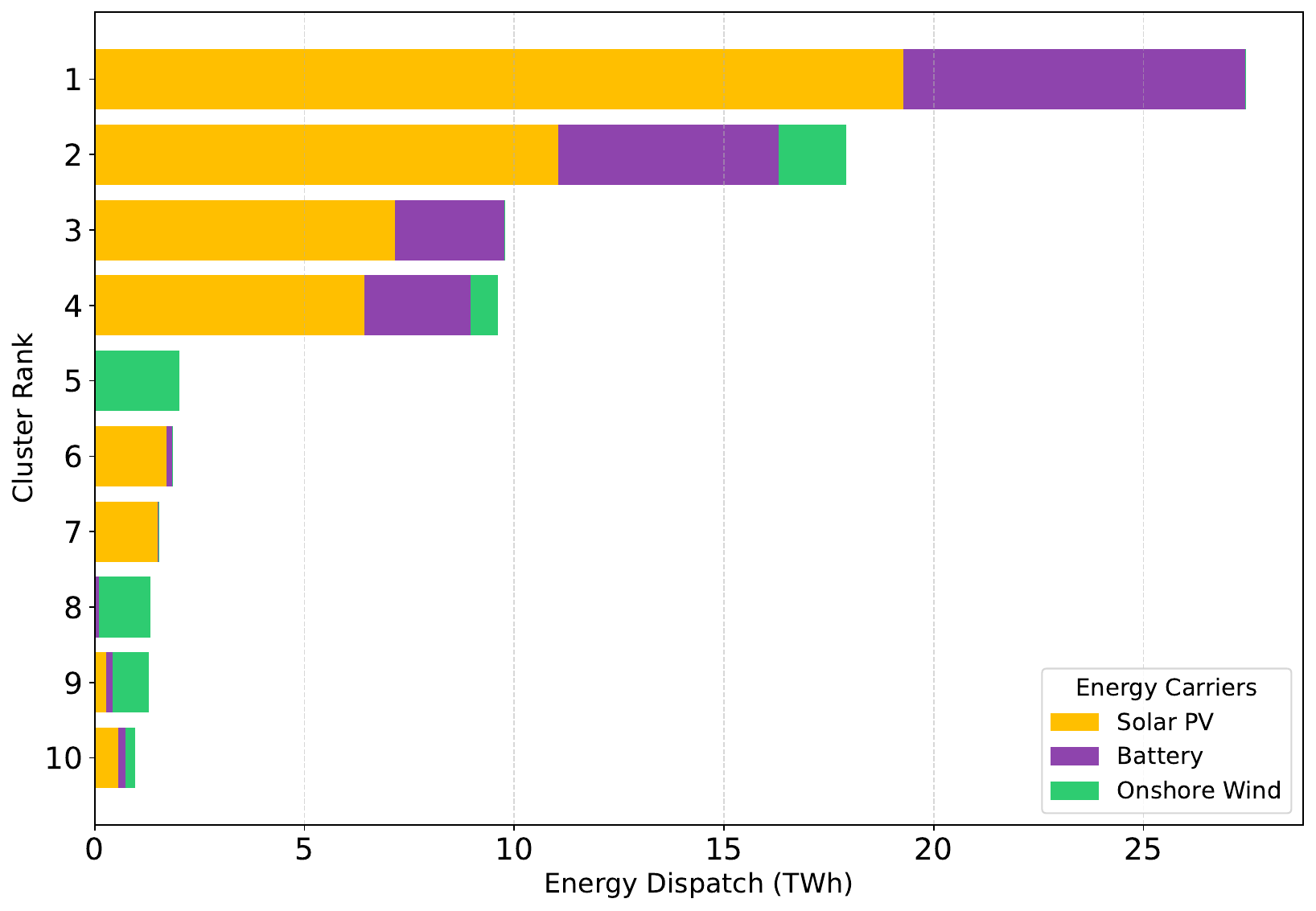}
    \caption{}
    \label{fig:LCO2_barchart}
\end{subfigure}
\caption{Optimal generation and transmission layout for the electricity-only model (Zero-CO2 case): (a) Energy mix and peak capacity loading of transmission lines across the clusters, (b) Distribution of the optimal generation and storage capacity mix for the top 10 clusters.}
\label{fig:LCO2_energy_mix}
\end{figure}

Under the zero-$\text{CO}_2$ scenario, the power system shifts from coastal natural gas plants (CCGT/OCGT) to a renewables-dominated supply architecture, as shown in figures~\ref{fig:No_LCO2_energy_mix} and~\ref{fig:LCO2_energy_mix}. High-irradiance southern inland nodes host massive utility solar PV installations, while coastal load centers like Tripoli and Benghazi integrate local wind generation and battery storage to handle diurnal variability and support grid stability.

\subsection{Transmission bottlenecks}
Utilizing the dual variables of the transmission constraints, we identify key corridors requiring reinforcement to facilitate 100\% renewable penetration.
As shown in Figure~\ref{fig:LCO2_piecharts_map}, several critical transmission lines reach the upper capacity threshold of 70\%, while others experience lower peak utilization. Because the optimization model enforces this 70\% limit as an operational ceiling across the entire year, lines exhibiting this peak loading hit the binding constraint limit, indicating that these corridors operate at their maximum allowed capacity during peak dispatch periods while remaining less loaded for the remainder of the year.

\subsection{System cost breakdown}
The overall annualized system cost increases by 31.4\% when transitioning from the unconstrained baseline to the carbon-limited scenario, rising from 3.75 to 4.93 billion~\texteuro/year. 

\begin{figure}[ht]
    \centering
    \includegraphics[width=0.70\textwidth]{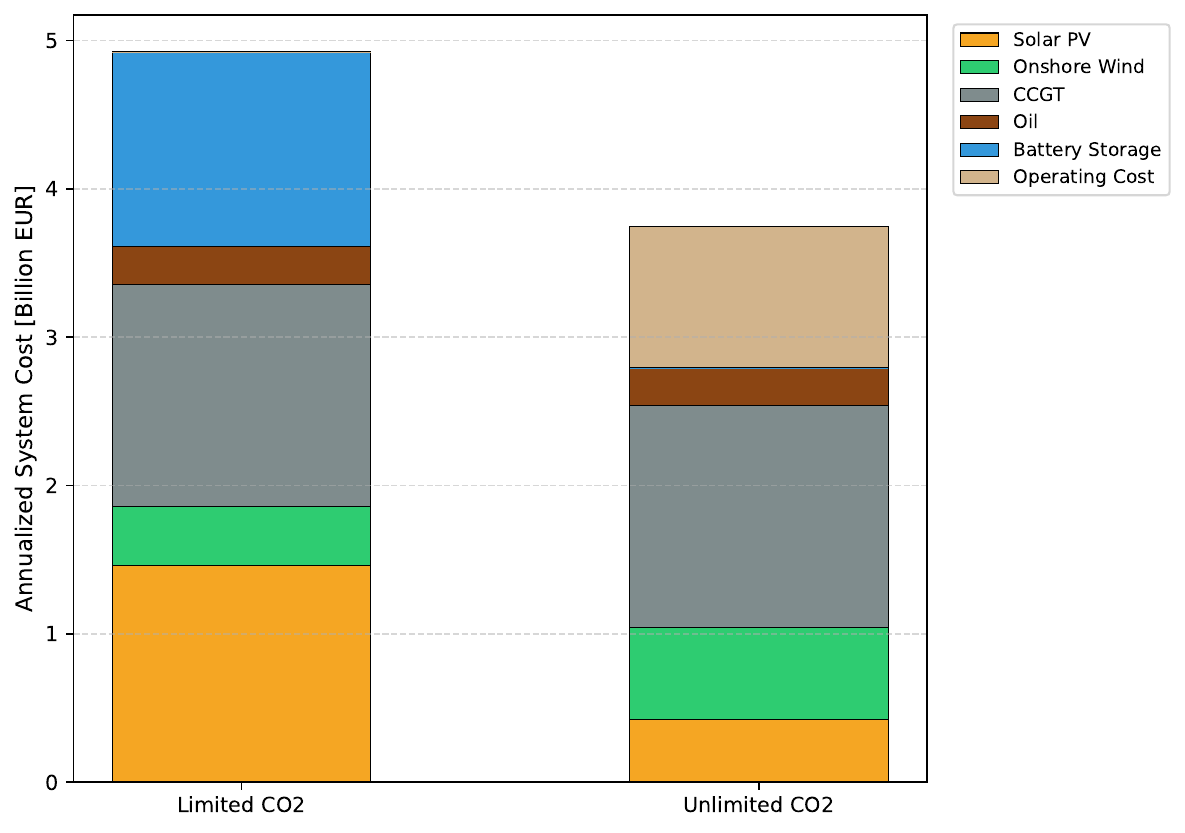}
    \caption{Annualized system cost breakdown by technology for the Limited $\text{CO}_2$ and Unlimited $\text{CO}_2$ electricity-only scenarios.}
    \label{fig:cost_breakdown}
\end{figure}

As shown in Figure~\ref{fig:cost_breakdown}, the primary driver behind this cost divergence is the capital expenditure required to firm up renewable generation under strict emission limits. In the Unlimited $\text{CO}_2$ case, fossil-fuel units are heavily dispatched to meet demand, incurring substantial operational and fuel costs (0.95 billion~\texteuro/year OPEX) while minimizing the need for energy storage CAPEX (0.001 billion~\texteuro/year). 

Conversely, enforcing emission constraints (Limited $\text{CO}_2$) shifts the system from a fuel-intensive operating model to a capital-intensive, solar-dominated paradigm:

\begin{itemize}
    \item {Solar PV capital expansion:} Investment in Solar PV capital expenditure (CAPEX) increases more than threefold, rising from 0.42 to 1.46 billion~\texteuro/year to exploit Libya's high solar irradiance. Conversely, Onshore Wind CAPEX decreases from 0.62 to 0.40 billion~\texteuro/year as Solar PV takes precedence.
    \item {Storage integration:} To bridge diurnal supply gaps without relying on high-emitting thermal dispatch, Battery Storage CAPEX surges from a negligible 0.001 billion~\texteuro/year to 1.3138 billion~\texteuro/year, accounting for 26.7\% of total system expenditure under the Limited $\text{CO}_2$ constraint.
    \item {Fossil displacement to reserve backup:} While CCGT and oil capital commitments (CAPEX) remain identical across both cases at 1.4972 and 0.2518 billion~\texteuro/year, respectively, their operational role switches from baseload/mid-merit generation to low-utilization reserve backup. Consequently, system-wide variable operating and fuel costs (OPEX) plummet to negligible levels (0.0013 billion~\texteuro/year).
\end{itemize}
\subsection{Financial analysis}
Enforcing zero $\text{CO}_2$ emissions requires a substantial upfront capital commitment to build out renewable energy generation and storage capacity. In contrast, the unconstrained, fossil-heavy system relies on significantly lower initial investment but incurs high ongoing operational and fuel expenses. 

To evaluate the economic trade-offs of full grid decarbonization in Libya, a financial feasibility comparison was conducted between the {Zero-$\text{CO}_2$ (Limited)} scenario and the unlimited baseline (unconstrained $\text{CO}_2$ emissions). Both configurations consider a 25-year operational horizon, a weighted average cost of capital (WACC) of 7\%, an annual demand of 53.65~TWh, and a benchmark electricity market price of 75~\texteuro/MWh. A summary of the financial parameters, annual cash flows, payback dynamics, and $\text{CO}_2$ emissions balance is presented in Table~\ref{tab:financial_co2_comparison}.

\begin{table}[htbp]
\centering
\begin{threeparttable}
\caption{Financial parameters, economic performance, and annual $\text{CO}_2$ emissions across the Zero-$\text{CO}_2$ and Unconstrained scenarios.}
\label{tab:financial_co2_comparison}
\small
\begin{tabular}{l r r}
\toprule
{Metric} & {Zero-$\text{CO}_2$ (Limited)} & {Unlimited} \\
\midrule
\multicolumn{3}{l}{\textit{{A. Financial Parameters \& System Costs}}} \\
Overnight CAPEX (\texteuro~B)\tnote{a} & 57.494 & 32.619 \\
Annualized CAPEX (\texteuro~B/yr) & 4.934 & 2.799 \\
Annual OPEX (\texteuro~M/yr) & 1.330 & 953.920 \\
Total Annual System Cost (\texteuro~B/yr) & 3.162 & 1.980 \\
{LCOE (\texteuro/MWh)} & {58.94} & {36.90} \\
\midrule
\multicolumn{3}{l}{\textit{{B. Profitability \& Valuation (at \texteuro~75/MWh Target Tariff)}}} \\
Annual Projected Revenue (\texteuro~B/yr)\tnote{b} & 4.024 & 4.024 \\
Annual Net Cash Flow (\texteuro~B/yr) & 4.022 & 3.070 \\
Net Present Value (NPV @ 7\%)\tnote{c} (\texteuro~B) & -10.617 & 3.157 \\
Internal Rate of Return (IRR) & 4.86\% & 8.05\% \\
Simple Payback Period (Years)\tnote{d} & 14.29 & 10.63 \\
Discounted Payback Period (Years)\tnote{e} & N/A & 20.12 \\
\midrule
\multicolumn{3}{l}{\textit{{C. Annual $\text{CO}_2$ Emissions Summary}}} \\
Total Annual Emissions ($\text{MtCO}_2$/yr) & 0.000 & 6.950 \\
System Carbon Intensity ($\text{g CO}_2$/kWh) & 0.00 & 0.13 \\
\bottomrule
\end{tabular}

\begin{tablenotes}
\footnotesize
\item[a] Overnight CAPEX represents the initial capital expenditure deployed at Year 0.
\item[b] Projected revenue is calculated using an assumed wholesale electricity price (target tariff) of 75~\texteuro/MWh applied to the total annual electricity demand of 53.65~TWh.
\item[c] NPV is calculated over a 25-year project horizon ($T = 25$) at a 7.0\% discount rate. It represents the single cumulative present value of all cash flows evaluated at Year 0: $\text{NPV} = -\text{CAPEX} + \sum_{t=1}^{25} \frac{\text{Revenue} - \text{OPEX}}{(1 + r)^t}$, where Year 0 reflects upfront capital outlay and Years 1--25 reflect discounted operational net cash flows.
\item[d] Simple Payback Period represents the undiscounted recovery timeframe in years, computed as $\text{Simple Payback} = \frac{\text{CAPEX}}{\text{Annual Net Cash Flow}}$.
\item[e] Discounted Payback Period (DPP) accounts for the time value of money at discount rate $r = 7\%$, computed analytically as $\text{DPP} = \frac{-\ln\left(1 - \frac{\text{CAPEX} \cdot r}{\text{Annual Net Cash Flow}}\right)}{\ln(1 + r)}$. For Zero-$\text{CO}_2$, payback is undefined (N/A) as discounted annual inflows do not cover annual capital costs at $r = 7\%$.
\end{tablenotes}
\end{threeparttable}
\end{table}

\subsection{Economic Takeaways and Policy Implications}
The financial and emission findings highlight several key structural insights regarding the energy transition in Libya:

\begin{itemize}
    \item {Front-loaded CAPEX vs. OPEX trade-off:} Achieving a zero-emission grid inflates overnight capital expenditures to 57.50~billion~\texteuro~(compared to 32.62~billion~\texteuro) due to extensive solar PV, wind, and battery storage investments. However, annual operational expenditures drop significantly to 1.33~M~\texteuro~(down from 953.92~million~\texteuro) as fossil fuel purchasing is virtually eliminated.
    \item {The ``Green premium'' and tariff discrepancy:} At the baseline electricity market price of 75~\texteuro/MWh, the unconstrained fossil system yields an LCOE of 36.90~\texteuro/MWh (resulting in a positive NPV of 3.157~billion~\texteuro, an Internal Rate of Return (IRR) of 8.05\%, and a payback period of 10.63 years). Conversely, full decarbonization drives the LCOE to 58.94~\texteuro/MWh (yielding an IRR of 4.86\% and extending the payback period to 14.29 years), establishing a green premium cost of 22.04~\texteuro/MWh.
    \item {Policy and financing requirements:} Although the Zero-$\text{CO}_2$ grid generates higher net annual operational cash flow (4.022~billion~\texteuro~vs.  3.070~billion~\texteuro), the extended payback period (14.29 years) and negative NPV (-10.617~billion~\texteuro) demonstrate that it fails to fully recover its massive initial investment under standard commercial financing (7\% WACC). To make zero-carbon infrastructure bankable (NPV $\ge 0$), targeted policy interventions are required—such as securing long-term wholesale power purchase agreements (PPAs) above 58.94 \texteuro/MWh, implementing carbon taxes on fossil fuels, or accessing concessional development financing with discount rates aligned with or below the project IRR of 4.86\%.
\end{itemize}

\section{Results: Sector-coupled hydrogen export} 
\label{sec:results_sector_coupled_h2_3secns} 
This section presents the results for a 70~TWh hydrogen export target across Scenarios A, B, and C, highlighting how alternative system configurations influence generation capacity, spatial investment patterns, and economic and environmental performance.

\subsection{Capacity expansion, infrastructure layout, and energy mix dispatch}
This subsection evaluates the spatial deployment of physical network infrastructure and optimized carrier dispatch across the three core scenarios. By contrasting regional transmission utilization alongside structural production footprints, it identifies the key geographical pivots driving the sector-coupled network optimization.

\begin{figure}[htbp]
\centering
\begin{subfigure}[b]{0.49\textwidth}
    \centering
    \includegraphics[height=5.2cm]{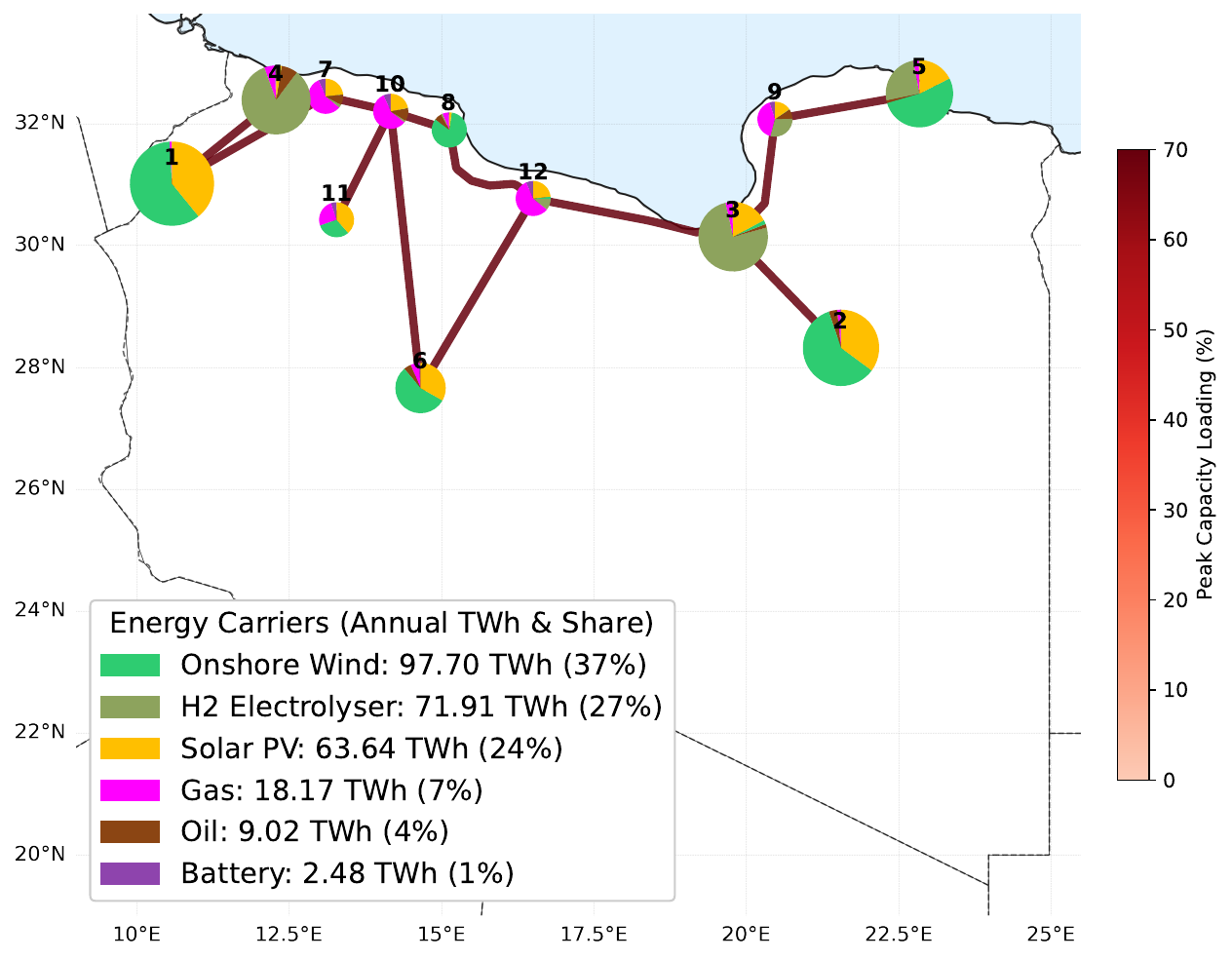}
    \caption{}
    \label{fig:Scenario_a_piecharts_map}
\end{subfigure}
\hfill
\begin{subfigure}[b]{0.49\textwidth}
    \centering
    \includegraphics[height=5.2cm]{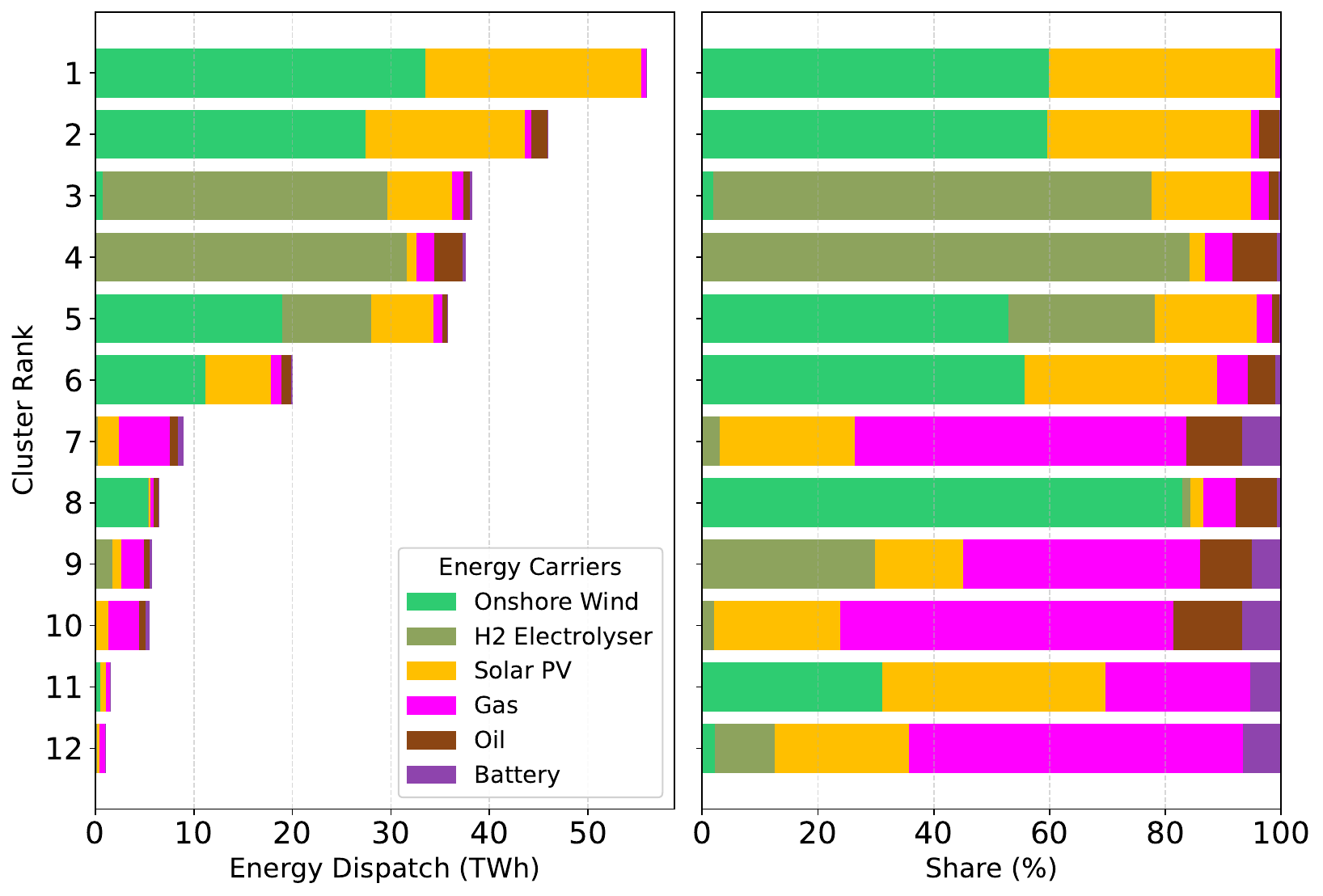}
    \caption{}
    \label{fig:Scenario_a_barchart}
\end{subfigure}
\caption{Spatial configuration of the optimal capacity expansion and transmission network under Scenario A: (a) Map of the spatial energy mix and peak capacity loading of transmission lines (\%) across the clusters, (b) Distribution of total annual energy dispatch across clusters, categorized by individual energy carriers and ranked by magnitude.}

\label{fig:Scenario_a_energy_mix}
\end{figure}

\begin{figure}[htbp]
\centering
\begin{subfigure}[b]{0.49\textwidth}
    \centering
    \includegraphics[height=5.2cm]{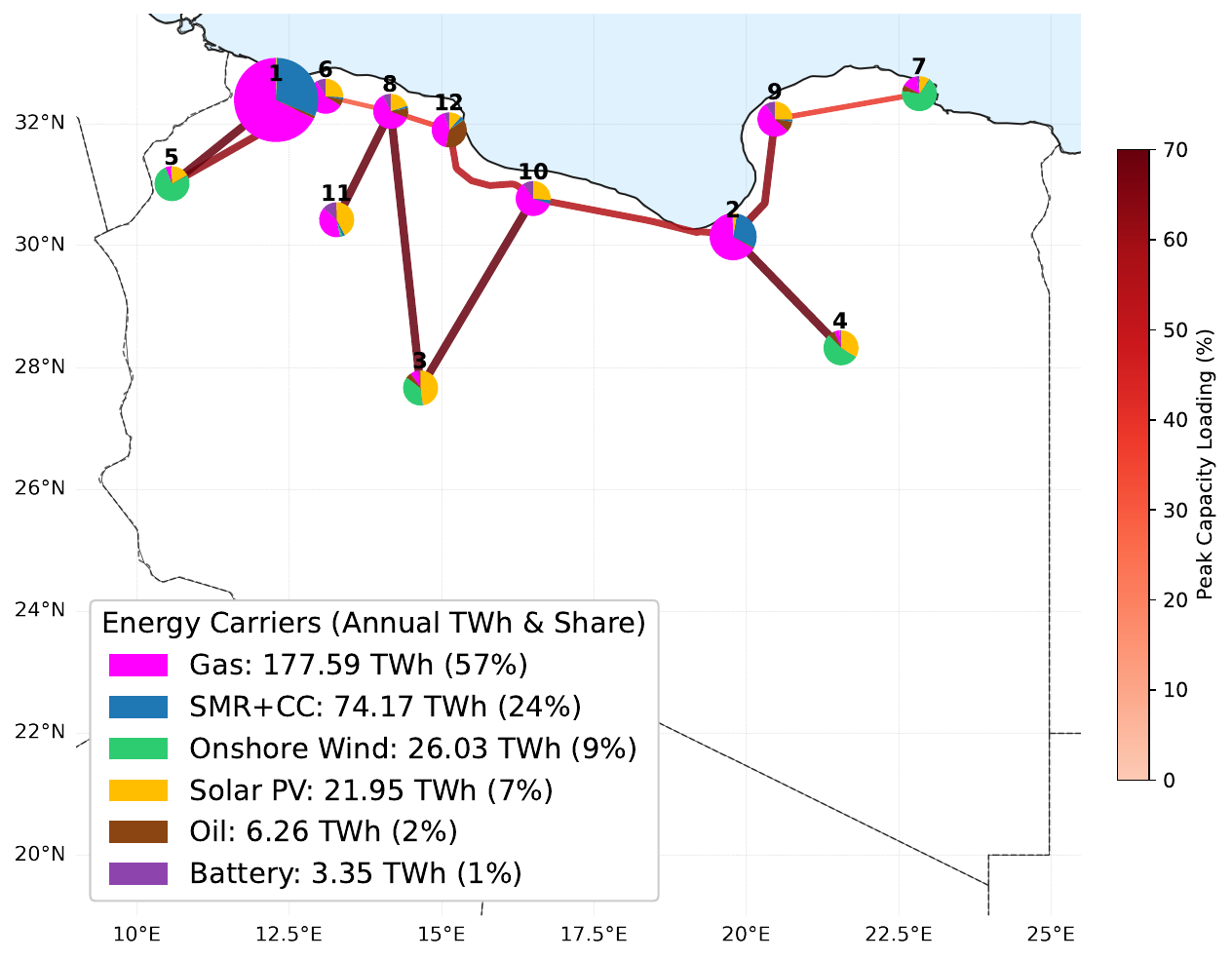}
    \caption{}
    \label{fig:Scenario_b_piecharts_map}
\end{subfigure}
\hfill
\begin{subfigure}[b]{0.49\textwidth}
    \centering
    \includegraphics[height=5.2cm]{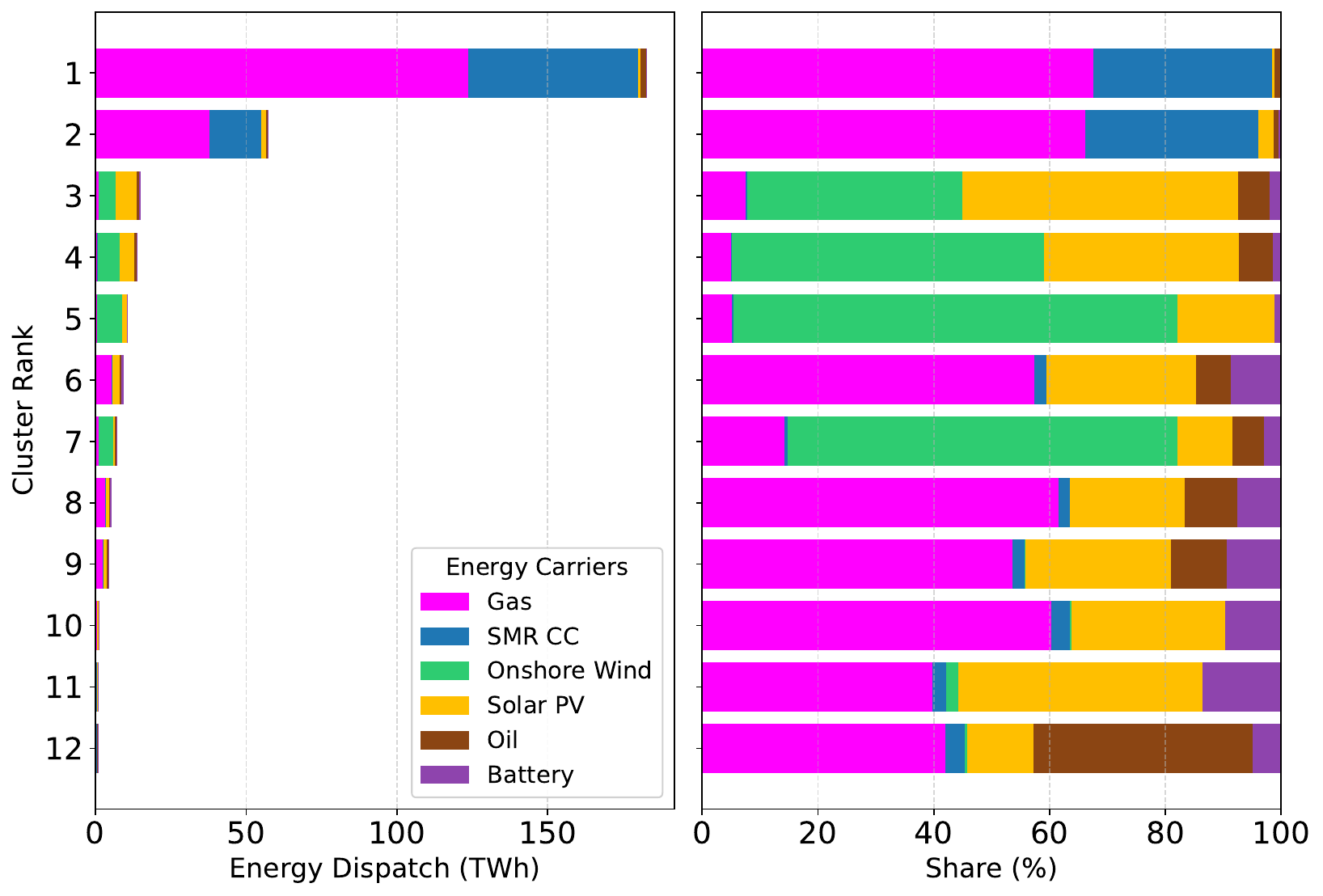}
    \caption{}
    \label{fig:Scenario_b_barchart}
\end{subfigure}
\caption{Spatial configuration of the optimal capacity expansion and transmission network under Scenario B: (a) Map of the spatial energy mix and peak capacity loading of transmission lines (\%) across the clusters, (b) Distribution of total annual energy dispatch across clusters, categorized by individual energy carriers and ranked by magnitude.}
\label{fig:Scenario_b_energy_mix}
\end{figure}

\begin{figure}[htbp]
\centering
\begin{subfigure}[b]{0.49\textwidth}
    \centering
    \includegraphics[height=5.2cm]{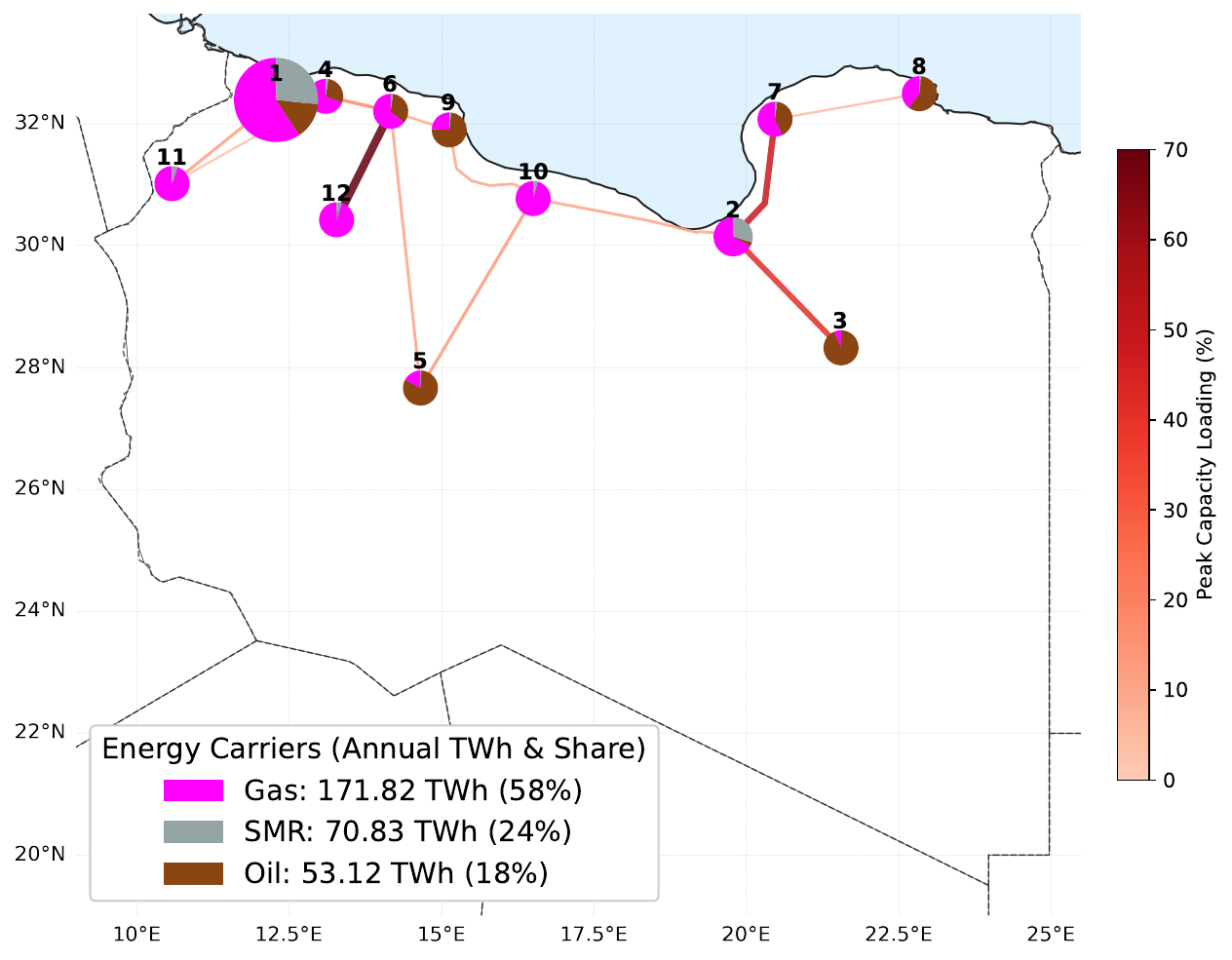}
    \caption{}
    \label{fig:Scenario_c_piecharts_map}
\end{subfigure}
\hfill
\begin{subfigure}[b]{0.49\textwidth}
    \centering
    \includegraphics[height=5.2cm]{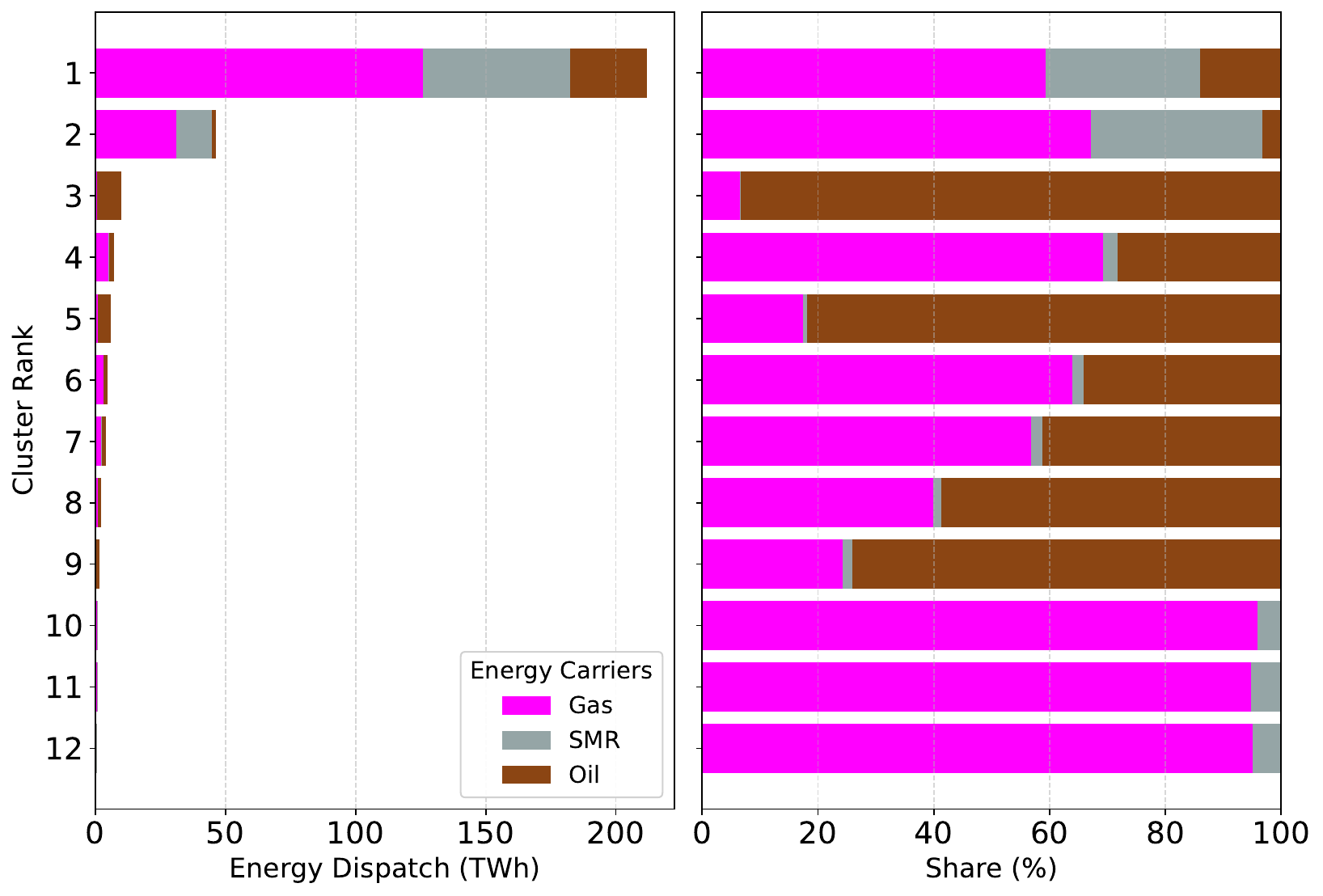}
    \caption{}
    \label{fig:Scenario_c_barchart}
\end{subfigure}
\caption{Spatial configuration of the optimal capacity expansion and transmission network under Scenario C: (a) Map of the spatial energy mix and peak capacity loading of transmission lines (\%) across the clusters, (b) Distribution of total annual energy dispatch across clusters, categorized by individual energy carriers and ranked by magnitude.}
\label{fig:Scenario_c_energy_mix}
\end{figure}

Figures~\ref{fig:Scenario_a_energy_mix}, \ref{fig:Scenario_b_energy_mix}, and \ref{fig:Scenario_c_energy_mix} illustrate the optimized carrier distribution and line loadings across Scenarios A, B, and C. A defining structural characteristic persistent across all three configurations is the complete dominance of Cluster 1 in the overall system profile.

As illustrated by the spatial layouts in figures~\ref{fig:Scenario_a_piecharts_map}, \ref{fig:Scenario_b_piecharts_map}, and \ref{fig:Scenario_c_piecharts_map}, hydrogen production capacity and supporting infrastructure are concentrated near high-irradiance inland areas and coastal export hubs. The optimizer strategically groups these assets to streamline supply logistics, efficiently bridging solar-rich inland generation paths with maritime export corridors. Furthermore, a comparison of transmission line loading reveals a clear gradient of network stress: congestion is highest and most widespread in Scenario A, decreases only modestly in Scenario B—where several key corridors still saturate at the 70\% limit—and drops sharply in Scenario C, reflecting significantly lower overall grid strain under that pathway's specific allocation.

Quantifying the energy distribution across clusters, the ranked cluster-level energy dispatch breakdowns in figures~\ref{fig:Scenario_a_barchart}, \ref{fig:Scenario_b_barchart}, and \ref{fig:Scenario_c_barchart}, energy activity is overwhelmingly concentrated at two main coastal gateways: Cluster 1 and Cluster 2. Cluster 1 represents the Al Zawia coastal region in the west, serving as the primary gateway for the Mellitah Oil \& Gas Complex, housing critical refinery infrastructure and deepwater maritime export facilities. Immediately following, Cluster 2 corresponds to Marsa Al Brega in the east, which acts as the central hub for eastern petrochemical processing, gas processing, and port operations. The combination of established industrial infrastructure, natural gas availability, and direct maritime port access at both hubs prompts the optimizer to co-locate blue hydrogen (SMR+CC) and energy dispatch at these two strategic coastal nodes, minimizing inland transmission bottlenecks and facilitating international export corridors. Concurrently, the remaining regional clusters function as significantly smaller nodes that primarily satisfy their own localized energy demands.

\subsection{System cost breakdown}
The total system expenditure exhibits notable variations across the evaluated scenarios, driven by the structural allocation of capital assets and operational expenses. To isolate the financial impact of large-scale hydrogen integration, annual expenditures are evaluated under both active export conditions and zero-export baseline conditions.

\begin{figure}[htbp]
\centering
\begin{subfigure}[b]{0.49\textwidth}
    \centering
    \includegraphics[width=\textwidth]{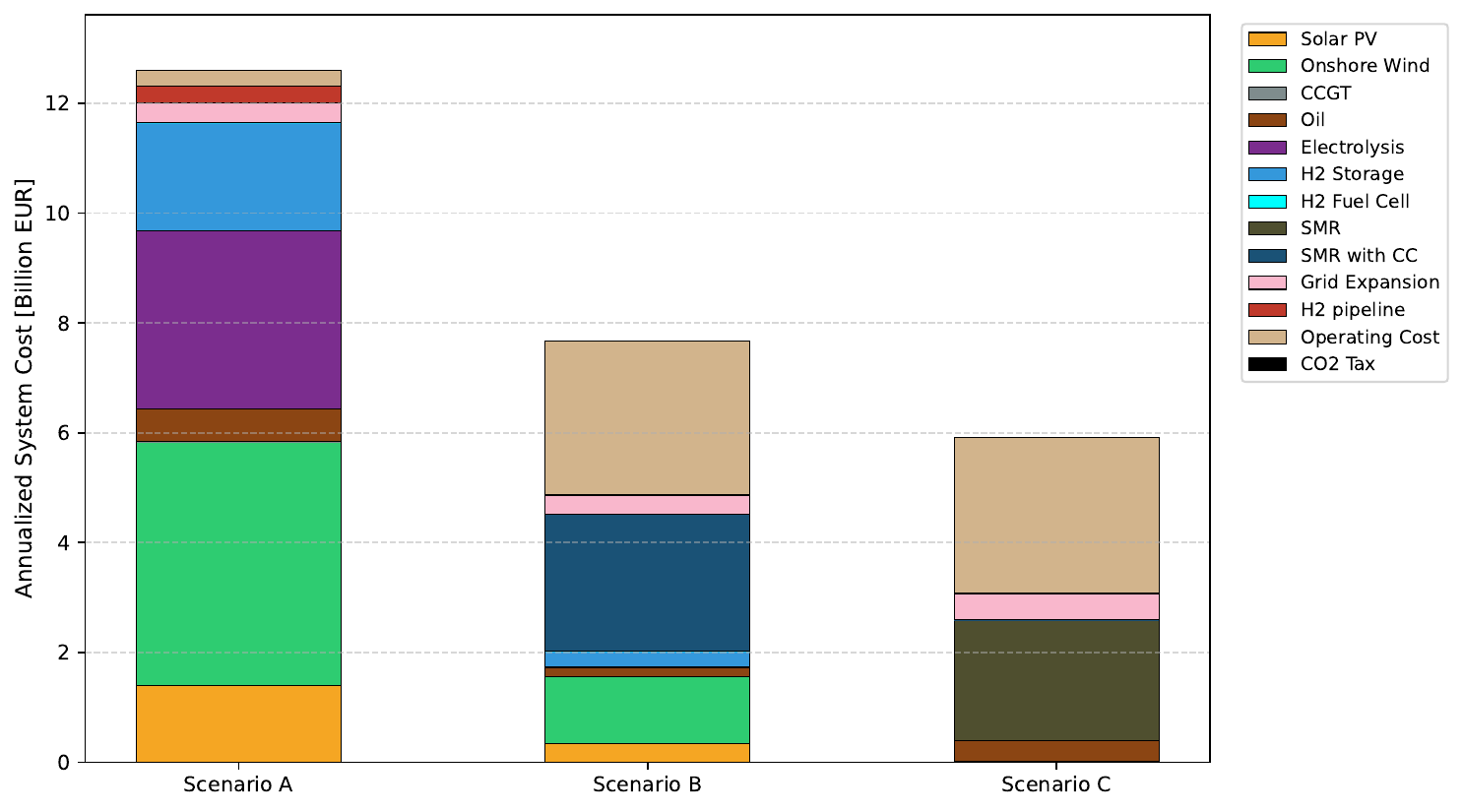}
    \caption{With hydrogen export (70~TWh)}
    \label{fig:inv_costs_H2_eport}
\end{subfigure}
\hfill
\begin{subfigure}[b]{0.49\textwidth}
    \centering
    \includegraphics[width=\textwidth]{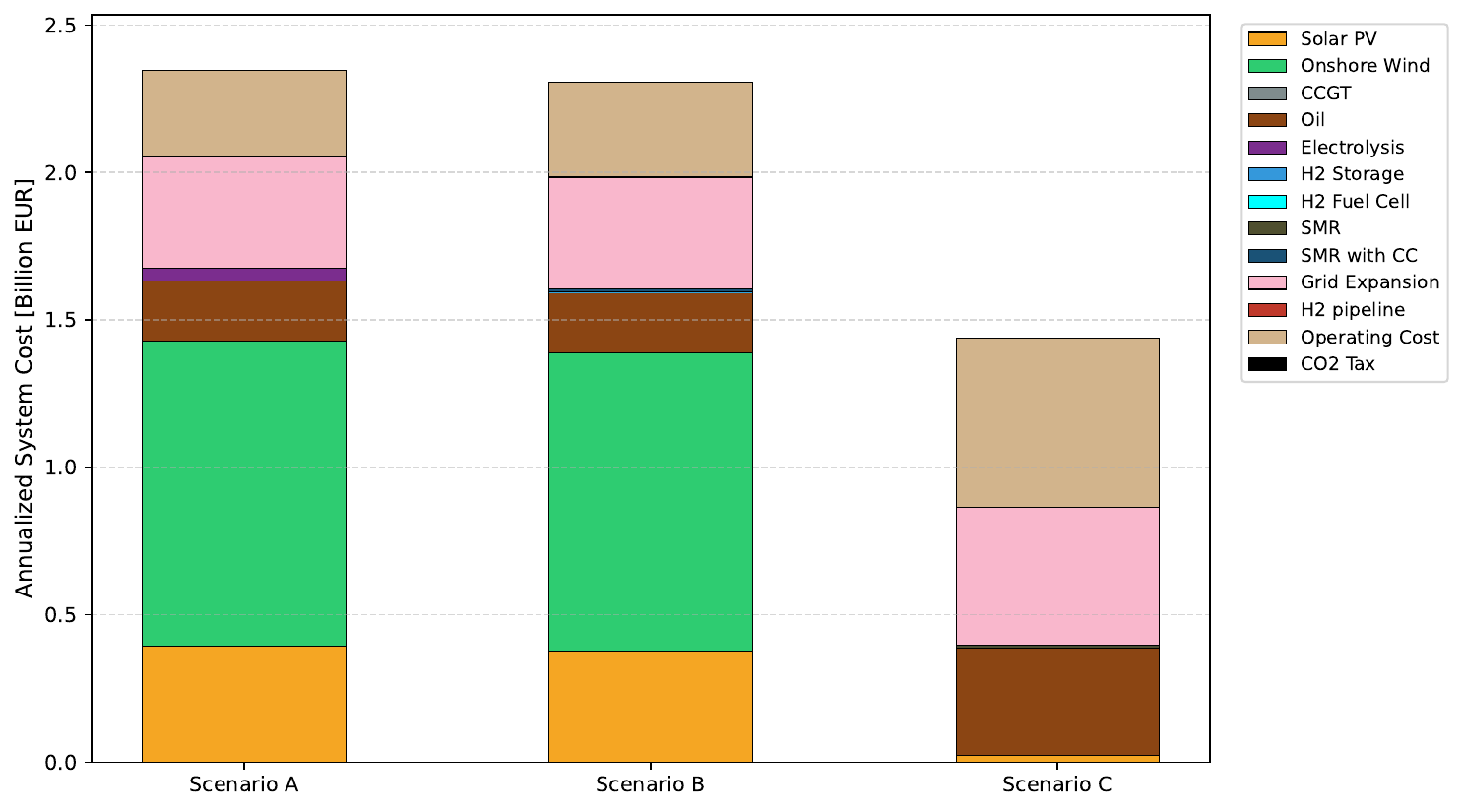}
    \caption{Zero-export baseline case}
    \label{fig:inv_costs_base}
\end{subfigure}
\caption{Total annual system cost comparison across the evaluated scenarios: (a) with hydrogen export at 70~TWh and (b) zero-export baseline case.}
\label{fig:annual_system_costs}
\end{figure}

Under the 70~TWh export demand (Fig.~\ref{fig:inv_costs_H2_eport}), total annual system expenditures range from 5.92 to 12.61~billion~\texteuro/yr. Scenario~A is heavily CAPEX-driven, dominated by dedicated renewable buildout (4.44~billion~\texteuro/yr Onshore Wind, 1.39~billion~\texteuro/yr Solar PV) and electrolysis infrastructure (3.23~billion~\texteuro/yr). In contrast, Scenarios~B and C shift financial burden toward operational expenditures (2.80 and 2.84~billion~\texteuro/yr, respectively), substituting massive wind/PV investments with grid-supplied electricity and Steam Methane Reforming with Carbon Capture (2.50~billion~\texteuro/yr in Scenario~B). Conversely, the zero-export baseline (Fig.~\ref{fig:inv_costs_base}) exhibits substantially lower overall system costs (1.44 to 2.35~billion~\texteuro/yr), where expenditures across all scenarios remain primarily tied to satisfying domestic electricity demand via modest renewable capacity, legacy oil generation, and minimal operational overhead.

\begin{figure}[htbp]
\centering
\begin{subfigure}[b]{0.49\textwidth}
    \centering
    \includegraphics[width=\textwidth]{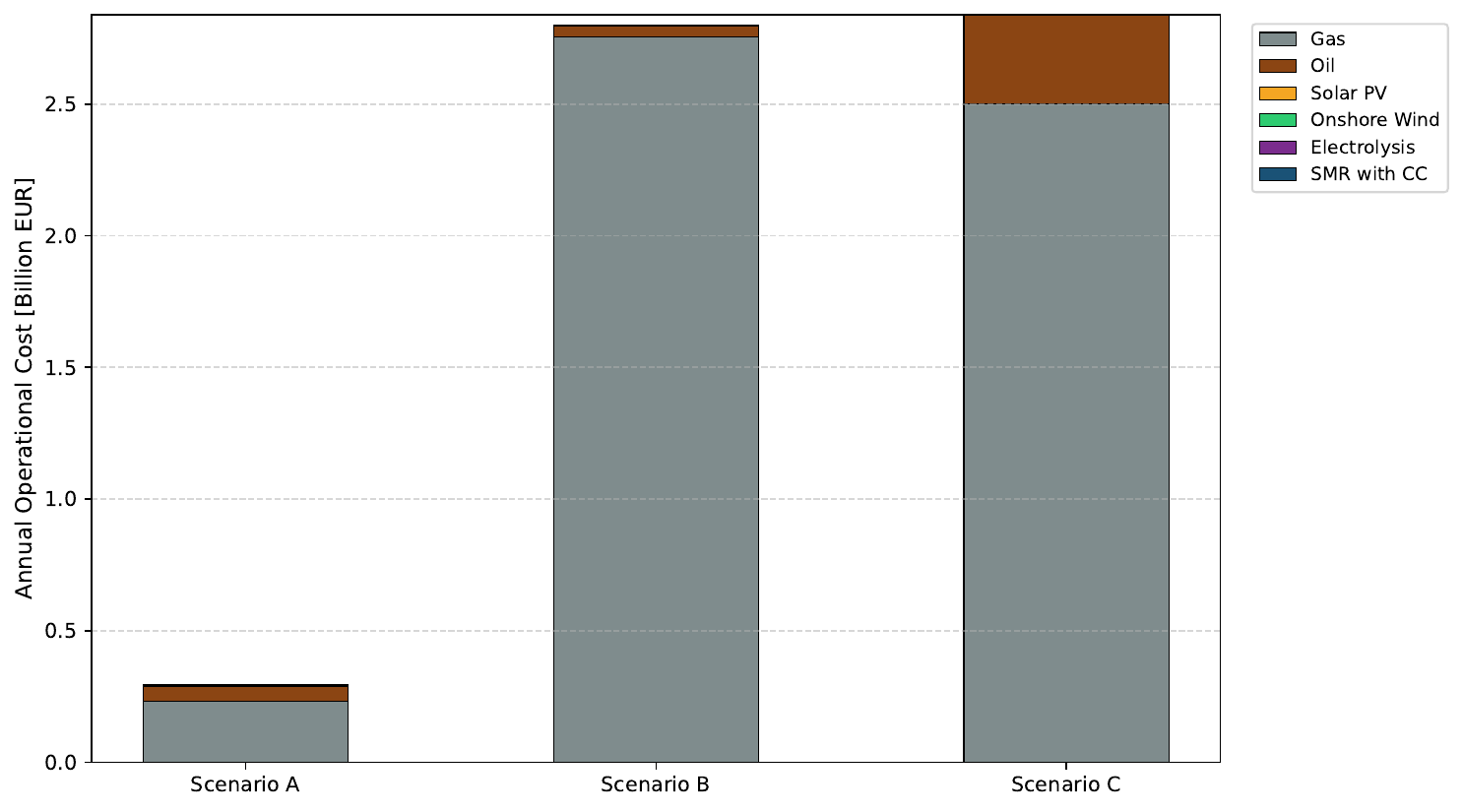}
    \caption{With hydrogen export (70~TWh)}
    \label{fig:opex_costs_H2_export}
\end{subfigure}
\hfill
\begin{subfigure}[b]{0.49\textwidth}
    \centering
    \includegraphics[width=\textwidth]{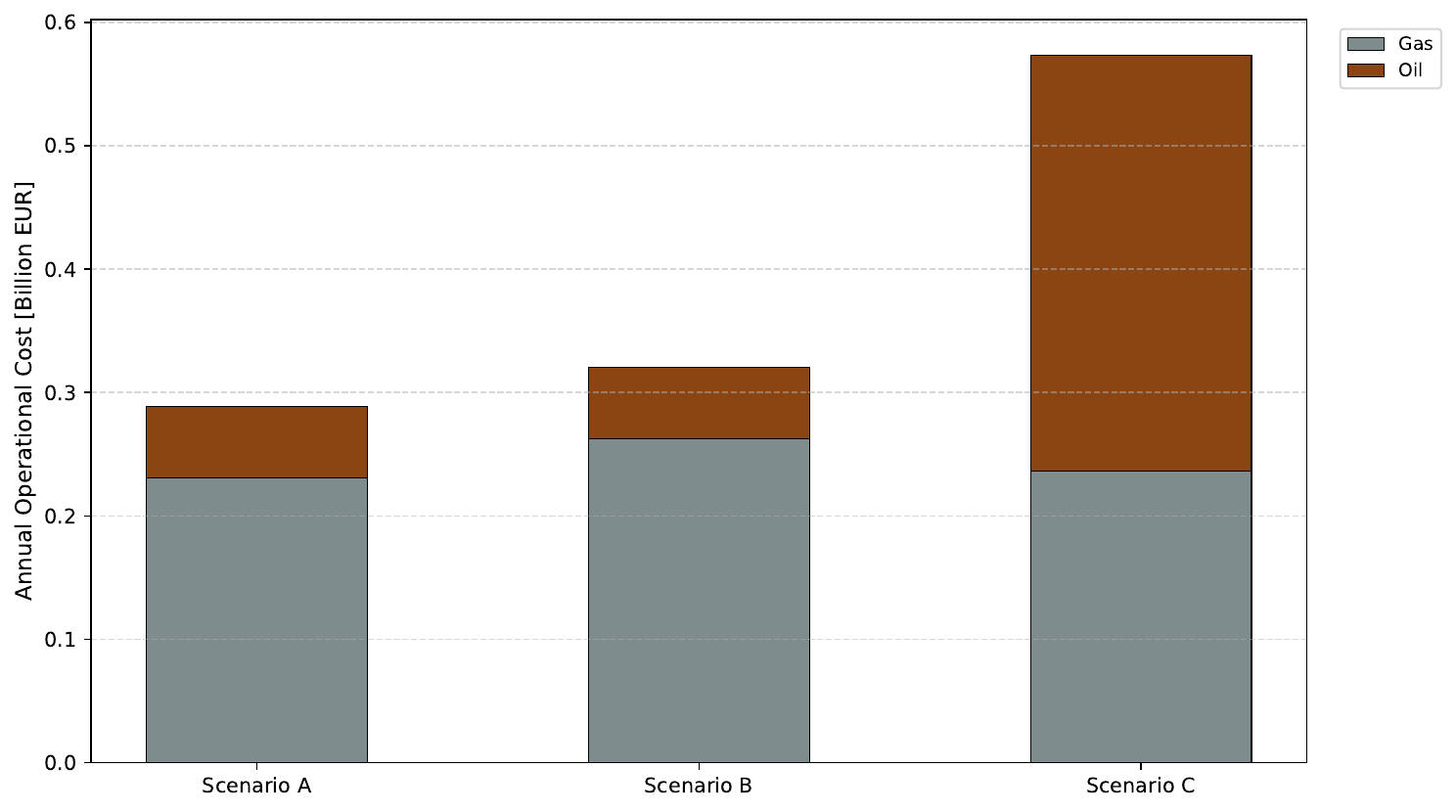}
    \caption{Zero-export baseline case}
    \label{fig:opex_costs_base}
\end{subfigure}
\caption{Operational Expenditure (OPEX) breakdown across the evaluated scenarios: (a) with hydrogen export at 70~TWh and (b) zero-export baseline case.}
\label{fig:annual_opex_costs}
\end{figure}

The OPEX breakdown (Fig.~\ref{fig:annual_opex_costs}) highlights a stark divergence between export and domestic-only cases. Under 70~TWh export demand (Fig.~\ref{fig:opex_costs_H2_export}), total OPEX surges from 0.30~billion~\texteuro/yr in Scenario~A to nearly 2.80 and 2.84~billion~\texteuro/yr in Scenarios~B and C, overwhelmingly driven by natural gas costs for SMR and thermal generation. Conversely, without exports (Fig.~\ref{fig:opex_costs_base}), system-wide OPEX remains below 0.60~billion~\texteuro/yr across all scenarios, driven predominantly by baseline legacy fuel demand. Notably, across all cases, primary fuel supply (gas and oil) overwhelmingly dominates operational expenditure, rendering non-fuel OPEX components negligible.

\subsection{Levelized cost of hydrogen (LCOH)}
To evaluate the economic viability of hydrogen production for export, we compute the net levelized cost of hydrogen across the considered scenarios. This metric provides a standardized basis for comparing production economics under varying grid integration strategies.

To isolate the incremental cost of hydrogen production from existing domestic power grid requirements, we utilize a paired-run subtraction method. For each scenario, the model is solved under the 70~TWh export demand (Fig.~\ref{fig:inv_costs_H2_eport}) and compared against a corresponding zero-export baseline (Fig.~\ref{fig:inv_costs_base}). The Net Levelized Cost of Hydrogen (Net LCOH) is defined as:
\begin{equation}
\label{Eq:Net-LCOH}
    \text{Net LCOH} = \frac{\text{Total Cost}_{\text{Export}} - \text{Total Cost}_{\text{Baseline}}}{\text{Exported Hydrogen Energy}}
\end{equation}
This approach ensures that domestic electricity consumers are not modeled as subsidizing the export infrastructure, providing a true marginal cost of hydrogen for the European market.

\begin{table}[htbp]
\centering
\caption{Summary of $\text{H}_2$ Export Costs and Net LCOH Breakdown}
\label{tab:h2_export_costs}
\begin{tabular}{lccc}
\toprule
Metric & Scenario A & Scenario B & Scenario C \\
\midrule
Export Run Cost [Billion \texteuro] & 12.61 & 7.67 & 5.91 \\
Baseline Zero-Export Cost [Billion \texteuro] & 2.35 & 2.31 & 1.44 \\
Net $\text{H}_2$ Cost [Billion \texteuro] & 10.26 & 5.36 & 4.48 \\
\midrule
Net LCOH [\texteuro/MWh] & 146.56 & 76.59 & 63.95 \\
Net LCOH LHV [\texteuro/kg] & 4.89 & 2.55 & 2.13 \\
Net LCOH HHV [\texteuro/kg] & 5.78 & 3.02 & 2.52 \\
\bottomrule
\multicolumn{4}{p{0.88\textwidth}}{\footnotesize \textit{Note:} Net $\text{H}_2$ Cost = Export System Cost -- Baseline Zero-Export Cost; Net LCOH = Net $\text{H}_2$ Cost / Exported $\text{H}_2$ Energy (70 TWh). Conversions assume $\text{H}_2$ Lower Heating Value (LHV) = $0.03333~\text{MWh/kg}$ and Higher Heating Value (HHV) = $0.03941~\text{MWh/kg}$.}
\end{tabular}
\end{table}

As summarized in Table~\ref{tab:h2_export_costs}, the Net LCOH exhibits significant divergence across the evaluated configurations. Across the 70~TWh export target, Scenario A yields the highest Net LCOH at 146.56~\texteuro/MWh (4.89~\texteuro/kg LHV), driven by the requirement for full renewable off-grid system expansion and dedicated asset over-provisioning. In contrast, Scenario B substantially reduces this metric to 76.59~\texteuro/MWh (2.55~\texteuro/kg LHV) through grid interaction and shared flexibility. Scenario C achieves the lowest overall production cost at 63.95~\texteuro/MWh (2.13~\texteuro/kg LHV), highlighting the strong economic advantage of co-locating generation and leveraging existing national transmission infrastructure.

\subsection{Net present value (NPV)}
Table~\ref{tab:npv_summary} presents the financial performance under a dynamic margin framework, applying a constant revenue premium of +€1.00/kg above the levelized cost of hydrogen (LCOH LHV) for each scenario over the 21-year operational horizon (2030–2050 at $r = 7.10\%$).
\begin{table}[ht]
\centering
\caption{Annual operational cash flows, project return metrics, and multi-year discounted financial evaluation by scenario (2030–2050) under a dynamic cost-plus revenue margin pricing model ($+\text{€}1.00/\text{kg}$ over scenario-specific LCOH at LHV) and an assumed zero $\text{CO}_2$ emission tax, evaluated at a discount rate of $r = 7.10\%$.}
\label{tab:npv_summary}
\small
\setlength{\tabcolsep}{8pt}
\resizebox{\linewidth}{!}{%
\begin{tabular}{lccc}
\hline
{Financial Metric} & {Scenario A} & {Scenario B} & {Scenario C} \\ \hline
LCOH LHV [\texteuro/kg] & 4.89 & 2.55 & {2.13} \\
Selling Price [\texteuro/kg] & 5.89 & 3.55 & {3.13} \\
Annual Cost [B~\texteuro/yr] & 12.61 & 7.67 & {5.91} \\
Annual Revenue [B~\texteuro/yr] & 12.36 & 7.46 & {6.58} \\
Annual Net Cash Flow [B~\texteuro/yr] & -0.25 & -0.21 & {+0.66} \\
Discounted Cost [B~\texteuro] & 135.51 & 82.43 & {63.58} \\
Discounted Revenue [B~\texteuro] & 132.85 & 80.20 & {70.69} \\
Total System NPV [B~\texteuro] & -2.65 & -2.23 & {+7.11} \\
Project IRR (\%) & 9.56\% & 15.27\% & {17.54\%} \\
Payback Period [Years] & 8.92 & 6.22 & {5.51} \\ \hline
\end{tabular}%
}
\end{table}

Under this pricing strategy, Scenario C demonstrates clear economic superiority by yielding a strong positive total system net present value (NPV). Achieving the lowest levelized cost of hydrogen at lower heating value (LCOH LHV) of €2.13/kg and a selling price of €3.13/kg, Scenario C delivers an annual revenue of €6.58~billion/yr against an annual system cost of €5.91~billion/yr, resulting in a positive annual net cash flow of +€0.66~billion/yr and a total system NPV of +€7.11~billion. Furthermore, Scenario C achieves an attractive Project IRR of 17.54\% and capital recovery with a payback period of 5.51 years, significantly outperforming Scenario B (15.27\% IRR; 6.22-year payback) and Scenario A (9.56\% IRR; 8.92-year payback).

By contrast, Scenarios A and B yield negative annual net cash flows (-€0.25~billion/yr and -€0.21~billion/yr) and negative total system NPVs (-€2.65~billion and -€2.23~billion, respectively) under the +€1.00/kg margin assumption due to their higher discounted system costs (€135.51~billion and €82.43~billion). Selling at €3.55/kg in Scenario B and €5.89/kg in Scenario A is insufficient to cover total system-wide expenditures over the 21-year period. Consequently, Scenario C remains the most financially resilient option, generating the strongest returns while keeping market-clearing hydrogen export prices competitive.

\subsection{Environmental evaluation}
In this section, the performance of the model is evaluated in terms of the environmental factors.

\subsubsection{CO$_2$ emissions}
\label{sec:co2_emis}
As detailed in Table~\ref{tab:co2_mass_balance}, gross $\text{CO}_2$ generation is partitioned into net atmospheric emissions (vented) and geologically sequestered emissions (captured via CCS). Scenarios A and B operate well within the carbon cap ceiling, resulting in identical net atmospheric emissions of 6.82~million tonnes (Mt) while relying on CCUS infrastructure to manage total system generation (capturing 25.22~Mt and 43.55~Mt, respectively). Conversely, the unconstrained baseline in Scenario C yields a significantly higher total gross generation of 135.04~Mt, consisting of 59.15~Mt in atmospheric venting and 75.90~Mt in carbon capture.

\begin{table}[htbp]
\centering
\caption{Gross $\text{CO}_2$ generation partitioned into net atmospheric emissions (vented) and geologically sequestered emissions (captured via CCS), reported in million tonnes (Mt).}
\label{tab:co2_mass_balance}
\begin{tabular}{lrrr}
\toprule
{$\text{CO}_2$ Accounting Metric} & {Scenario A} & {Scenario B} & {Scenario C} \\ \midrule
Net Atmospheric $\text{CO}_2$ Emissions (Vented) & 6.82 Mt & 6.82 Mt & 59.15 Mt \\
Geologically Sequestered $\text{CO}_2$ (Captured) & 25.22 Mt & 43.55 Mt & 75.90 Mt \\ \midrule
Total Gross $\text{CO}_2$ Generated (Vented + Stored) & 32.05 Mt & 50.37 Mt & 135.04 Mt \\ \bottomrule
\end{tabular}
\end{table}

\subsubsection{Water consumption}
\label{sec:desalination}
A critical factor in the feasibility of large-scale hydrogen production in arid regions like Libya is the required water supply. In this study, the costs and energy overhead associated with Reverse Osmosis (RO) desalination are integrated directly into the hydrogen supply chain assumptions, accounting for the technology-specific water intensities of electrolysis, SMR with carbon capture, and conventional SMR.

Our analysis indicates that total hydrogen output ranges between $2.13\text{--}2.16~\text{Mt}$ ($70.91\text{--}71.91~\text{TWh}_{\text{H2}}$) across the evaluated scenarios. However, water demand varies significantly based on the technology mix deployed, requiring $21.57~\text{Mm}^3$ in Scenario A, $15.96~\text{Mm}^3$ in Scenario B, and $10.65~\text{Mm}^3$ in Scenario C (Table~\ref{tab:water_demand}).

\begin{table}[htbp]
\centering
\caption{Estimated water demand and desalination costs across scenarios.}
\label{tab:water_demand}
\begin{tabular}{lccc}
\toprule
Metric & Scenario A & Scenario B & Scenario C \\
\midrule
$\text{H}_2$ Produced ($\text{TWh}_{\text{H2}}$) & 71.91 & 70.92 & 70.91 \\
$\text{H}_2$ Mass (Million Tonnes) & 2.157 & 2.128 & 2.128 \\
Water Demand (Million $\text{m}^3$) & 21.57 & 15.96 & 10.65 \\
Est. Desalination Cost (\text{M\texteuro/yr}) & 32.36 & 23.94 & 15.97 \\
\bottomrule
\end{tabular}
\end{table}

Despite the scale of these volumes, the economic impact of water supply is remarkably low. The estimated desalination costs contribute between 1.04\% and 1.38\% to the total annualized system cost, representing a marginal addition of only €0.0075 to €0.0150 per kg of hydrogen (€0.0075--0.0150/kg). 

From an energy perspective, the overhead required for RO desalination results in a specific energy consumption of just $0.0175\text{--}0.0350~\text{kWh/kg}_{\text{H2}}$. When evaluated against total system electricity demand, desalination accounts for merely $0.033\%\text{--}0.067\%$. To prevent the depletion of fragile inland fossil aquifers in hyper-arid regions like Ghadames, green hydrogen production in Scenario A is spatially restricted to coastal clusters, ensuring direct integration with seawater reverse-osmosis (SWRO) facilities. These findings confirm that while water sourcing requires geographical planning (e.g., coastal co-location of RO plants), water consumption does not act as an economic or energetic bottleneck for Libya's hydrogen supply potential.

\subsection{System cost dynamics under $\text{CO}_2$ taxation}
The inclusion of a $100\text{~\texteuro/tonne}$ $\text{CO}_2$ tax fundamentally alters system cost dynamics, shifting economic penalties away from capital expenditure and toward carbon-intensive operation.

\begin{figure}[htbp]
    \centering
    \begin{subfigure}[b]{0.49\textwidth}
        \centering
        \includegraphics[width=\textwidth]{Figures/fig_cost_by_scenario.pdf}
        \caption{Without carbon tax}
        \label{fig:cost_no_tax}
    \end{subfigure}
    \hfill
    \begin{subfigure}[b]{0.49\textwidth}
        \centering
        \includegraphics[width=\textwidth]{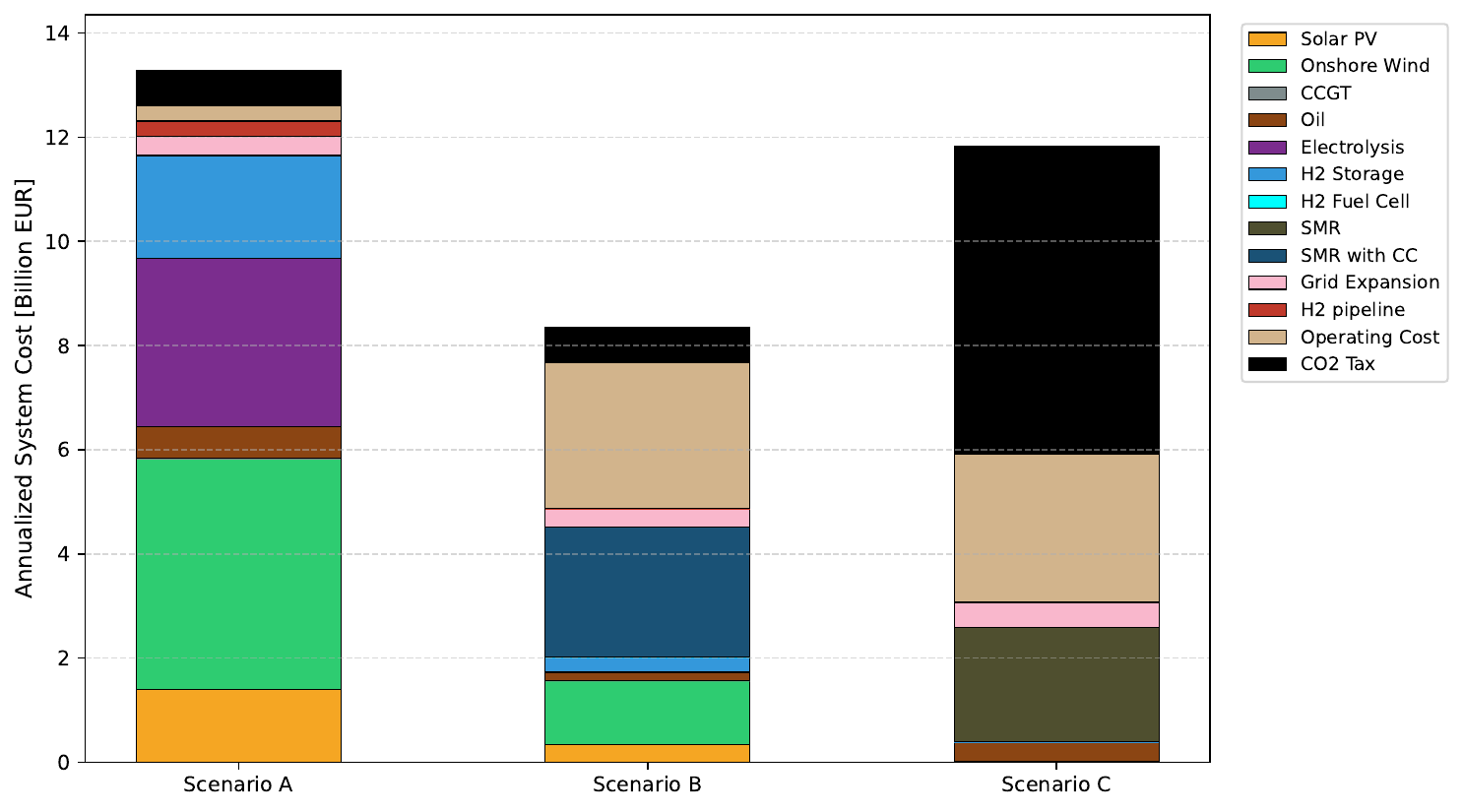}
        \caption{With $100\text{~\texteuro/tonne}$ $\text{CO}_2$ tax}
        \label{fig:cost_with_tax}
    \end{subfigure}
    \caption{Annualized total system cost breakdown across hydrogen export scenarios ($70\text{~TWh}$) under (a) untaxed conditions and (b) a $100\text{~\texteuro/tonne}$ $\text{CO}_2$ carbon tax.}
    \label{fig:cost_breakdown_CO2_Tax}
\end{figure}

The impact of carbon pricing on total system costs across export scenarios is illustrated in Figure~\ref{fig:cost_breakdown_CO2_Tax}.

\begin{itemize}
    \item {Tax Liability vs. CAPEX Expansion:} Unabated fossil reliance drives prohibitive penalty costs under carbon pricing. In Scenario~C, the CO$_2$ tax alone reaches 5.91~Billion~\texteuro/year, constituting $50.0\%$ of its total annual expenditure (11.83~Billion~\texteuro/year). Conversely, higher capital investments (\text{CAPEX}) in renewables and electrolysis (Scenario~A: 1.39~Billion~\texteuro/year Solar PV, 4.44~Billion~\texteuro/year Wind, 3.23\text{~Billion \texteuro/year} Electrolysis) limit carbon tax liabilities to just 0.68~Billion~\texteuro/year, increasing total system expenditure modestly from 12.61~Billion~\texteuro/year to 13.29~Billion~\texteuro/year.
    \item {Technology Competitiveness:} The 100~\texteuro/tonne threshold erodes the cost advantage of unabated Steam Methane Reforming (SMR). Scenario~B demonstrates that integrating Carbon Capture (SMR-CC) yields the lowest overall system cost under taxation (8.35~Billion~\texteuro/year), balancing capital investment with a heavily reduced carbon tax penalty (0.68~Billion~\texteuro/year).
    \item {Expenditure Composition:} Under carbon pricing, high-carbon pathways (Scenario~C) transition to an operational-liability model dominated by tax payments, increasing total costs from 5.92~Billion~\texteuro/year to 11.83~Billion~\texteuro/year, whereas decarbonized options shift system value toward fixed, asset-driven capital expenditure.
\end{itemize}

\subsection{Financial viability under carbon taxation}
A comparative summary of the core financial metrics under untaxed conditions ($0\text{~\texteuro/tCO}_2$) versus a $100\text{~\texteuro/tonne}$ $\text{CO}_2$ tax regime is presented in Table~\ref{tab:co2_tax_financial_comparison}.

\begin{table}[ht]
\centering
\small
\resizebox{\linewidth}{!}{%
\begin{threeparttable}
\caption{Comparison of key financial performance metrics across scenarios under zero tax versus a $100\text{~\texteuro/tonne}$ $\text{CO}_2$ tax ($r = 7.10\%$, 2030--2050 horizon, revenue margin $+1\text{~\texteuro/kg}$).}
\label{tab:co2_tax_financial_comparison}
\setlength{\tabcolsep}{6pt}
\begin{tabular}{lcccccc}
\toprule
& \multicolumn{2}{c}{Scenario A} & \multicolumn{2}{c}{Scenario B} & \multicolumn{2}{c}{Scenario C} \\ 
\cmidrule(lr){2-3} \cmidrule(lr){4-5} \cmidrule(lr){6-7}
Financial Metric & No Tax & Taxed & No Tax & Taxed & No Tax & Taxed \\ 
\midrule
Annual System Cost [B~\texteuro/yr] & 12.61 & 13.29 & 7.67 & 8.35 & 5.91 & 11.83 \\
Total System NPV [B~\texteuro] & -2.65 & -9.99 & -2.23 & -9.56 & +7.11 & -56.47 \\
Project IRR (\%) & 9.56\% & 8.78\% & 15.27\% & 12.77\% & 17.54\% & N/A\tnote{*} \\
Simple Payback Period [Years]\tnote{a} & 8.92 & 9.44 & 6.22 & 7.21 & 5.51 & N/A\tnote{*} \\ 
Discounted Payback Period [Years]\tnote{b} & 14.12 & 15.35 & 8.84 & 10.65 & 7.39 & N/A\tnote{*} \\ 
\bottomrule
\end{tabular}

\begin{tablenotes}
\footnotesize
\item[a] Simple Payback Period represents the undiscounted capital recovery timeframe in years, calculated as $\text{Simple Payback} = \frac{\text{Net Overnight CAPEX}}{\text{Annual Net Operating Cash Flow}}$.
\item[b] Discounted Payback Period (DPP) accounts for the time value of money ($r = 7.10\%$), evaluated analytically as $\text{DPP} = \frac{-\ln\left(1 - \frac{\text{CAPEX} \cdot r}{\text{Annual Net Cash Flow}}\right)}{\ln(1 + r)}$.
\item[*] N/A indicates that net annual operating cash flows are strictly negative (or annual discounted inflows are insufficient to recover capital costs at $r = 7.10\%$), rendering payback unattainable and IRR non-viable.
\end{tablenotes}
\end{threeparttable}%
}
\end{table}

The introduction of carbon pricing severely undermines project bankability across all pathways, though to vastly different degrees:

\begin{itemize}
    \item {Catastrophic Impact on Fossil Pathways:} Scenario~C experiences total economic collapse under carbon pricing. The 5.91~Billion~\texteuro/year tax liability drives total annual system costs up from 5.91~Billion~\texteuro/year to 11.83~Billion~\texteuro/year, rendering the total system NPV deeply negative at $-56.47$~Billion~\texteuro~(down from $+7.11$~Billion~\texteuro). Because annual net operating cash flows fall to $-5.25$~Billion~\texteuro/year, the system never reaches a financial breakeven point, yielding non-viable (N/A) figures for IRR and Payback Period.
    \item {Resilience of Decarbonized Pathways:} Scenarios~A and B absorb the tax policy far more effectively due to minimal residual emissions. While both experience total system NPV contractions (Scenario~A moving from $-2.65$~B~\texteuro~to $-9.99$~B~\texteuro; Scenario~B from $-2.23$~B~\texteuro~to $-9.56$~B~\texteuro), they preserve strong asset performance and positive net operational cash flows. Scenario~B retains a solid financial profile with a $12.77\%$ Project IRR and a payback period of 7.21~years.
\end{itemize}

\section{Results: Hybrid pathway for hydrogen export}
\label{sec:results_hybrid_h2} 
This section examines the results obtained under the hybrid scenario, which combines a 70~TWh hydrogen export target with a $100\text{~\texteuro/tonne}$ $\text{CO}_2$ carbon tax. The analysis focuses on the resulting generation capacity, spatial capacity allocation, and system costs under this combined policy and export setting.

\subsection{Capacity expansion, infrastructure layout, and energy mix dispatch}
This subsection details the spatial deployment of physical infrastructure and operational energy flows optimized for the hybrid scenario. By mapping transmission line utilization alongside localized supply profiles, it provides insights into how infrastructure grids co-optimize with regional resource variations to meet export demands.

\begin{figure}[htbp]
\centering
\begin{subfigure}[b]{0.49\textwidth}
    \centering
    \includegraphics[height=5.2cm]{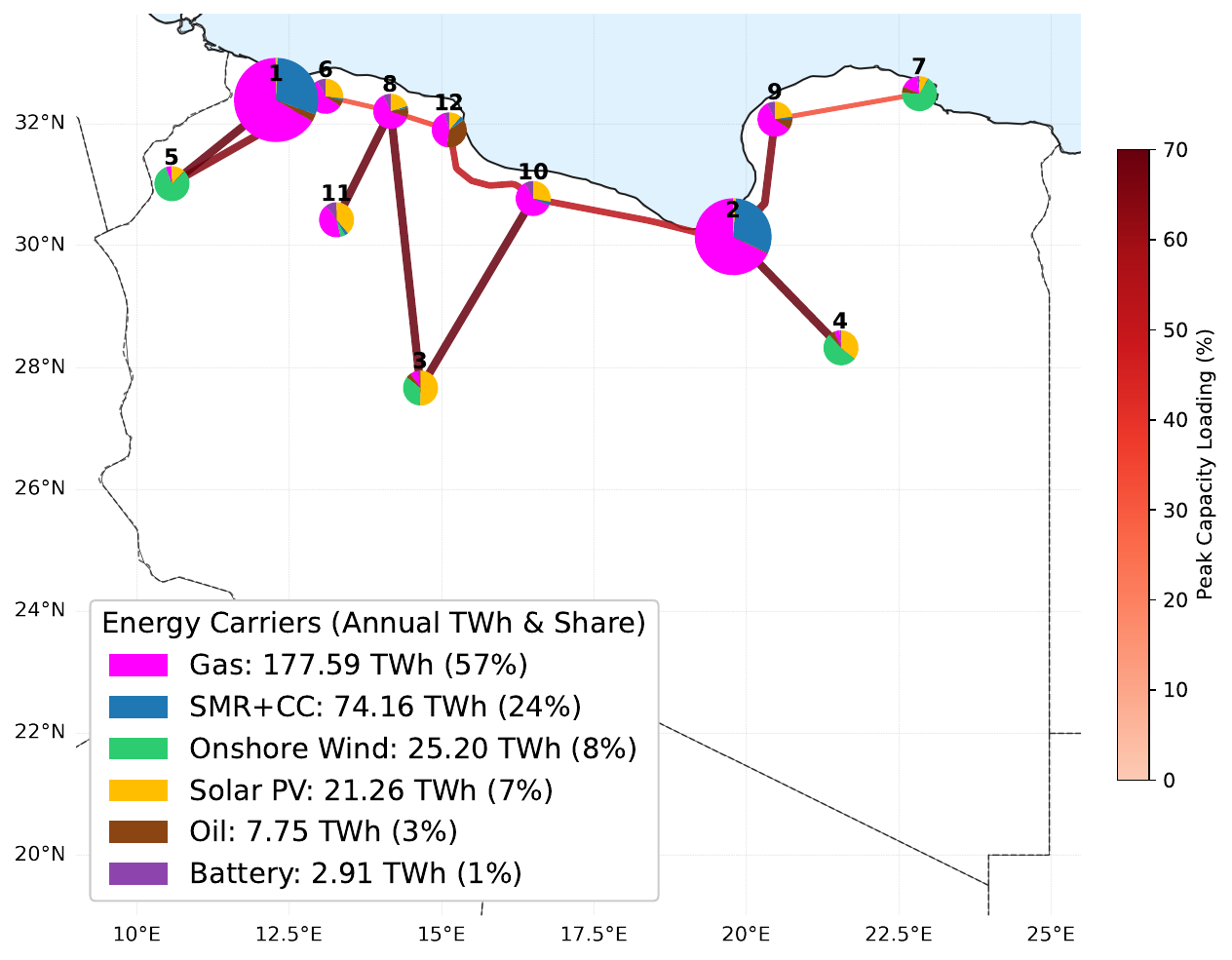}
    \caption{}
    \label{fig:Scenario_hybrid_piecharts_map}
\end{subfigure}
\hfill
\begin{subfigure}[b]{0.49\textwidth}
    \centering
    \includegraphics[height=5.2cm]{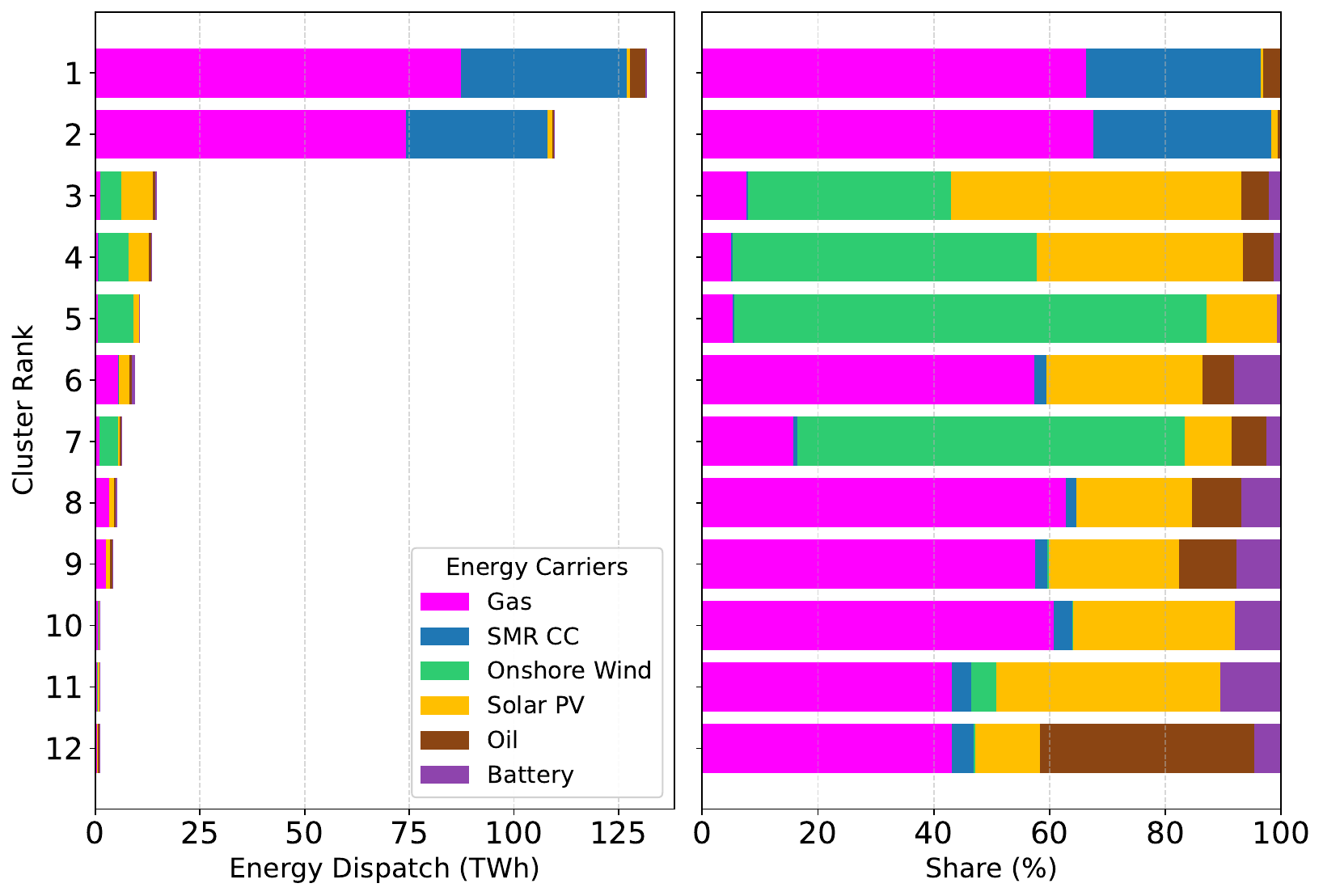}
    \caption{}
    \label{fig:Scenario_hybrid_barchart}
\end{subfigure}
\caption{Spatial configuration of the optimal capacity expansion and transmission network under hybrid scenario: (a) Map of the spatial energy mix and peak capacity loading of transmission lines (\%) across the clusters, (b) Distribution of total annual energy dispatch across clusters, categorized by individual energy carriers and ranked by magnitude.}
\label{fig:Scenario_hybrid_energy_mix}
\end{figure}

Figure~\ref{fig:Scenario_hybrid_energy_mix}a illustrates the geospatial layout of this carrier mix alongside peak transmission line capacity loading. The optimization heavily clusters system activity at the primary coastal export hubs—most notably at Cluster 1 (Al Zawia) in the west and Cluster 2 (Marsa Al Brega) in the east, while clean generation sources like onshore wind and solar PV are strategically deployed across the higher-irradiance internal nodes.

The absolute volumetric distribution across the network is captured in figure~\ref{fig:Scenario_hybrid_energy_mix}b, which ranks all 12 clusters by total energy dispatch magnitude. Cluster 1 represents the Al Zawia coastal region, serving as the strategic entry point for the Mellitah Oil \& Gas Complex, while Cluster 2 corresponds to Marsa Al Brega, the key eastern industrial hub for natural gas processing and petrochemical export. Because of this strategic infrastructure hubbing, energy dispatch across the network is heavily concentrated at these two dominant coastal clusters through a mix composed primarily of natural gas generation and SMR+CC facilities. Co-locating fuel supplies with chemical processing and deepwater maritime access eliminates inland transmission penalties, while the remaining internal clusters generate significantly smaller energy volumes to supply their localized demands.

\subsection{System cost breakdown}
Integrating a combined policy framework, pairing a $70\text{~TWh}$ hydrogen export mandate with a $100\text{~\texteuro/tonne}$ $\text{CO}_2$ carbon tax, reshapes the overall system economics by striking a balance between capital expenditure (\text{CAPEX}) investments and operational emission penalties.

\begin{figure}[htbp]
    \centering
    \begin{subfigure}[b]{0.40\textwidth}
        \centering
        \includegraphics[width=\textwidth]{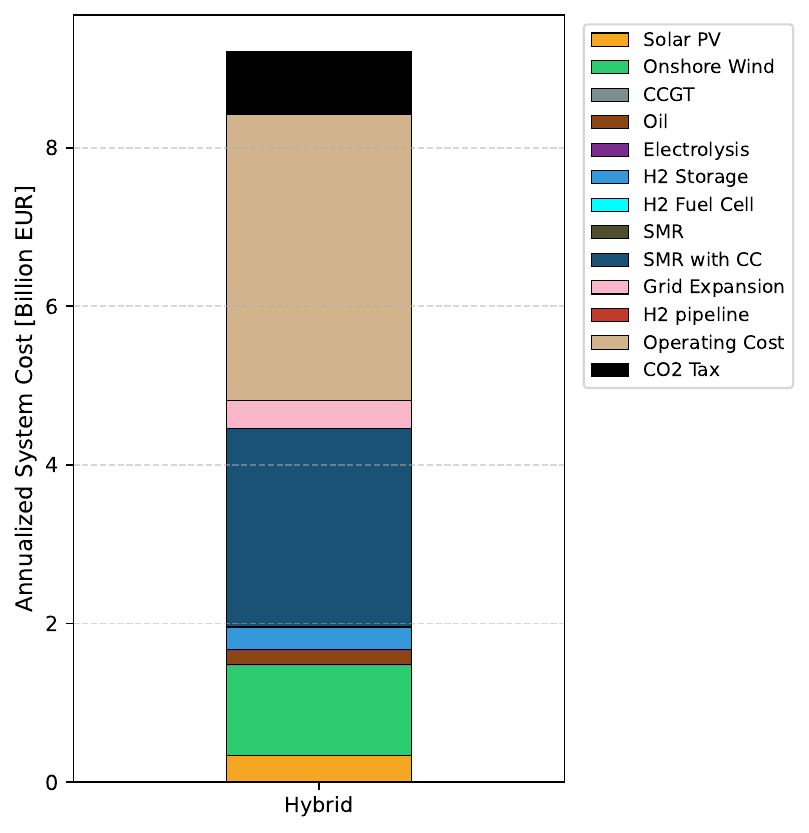}
        \caption{With hydrogen export (70~TWh)}
        \label{fig:cost_hybrid_export}
    \end{subfigure}
    \hfill
    \begin{subfigure}[b]{0.40\textwidth}
        \centering
        \includegraphics[width=\textwidth]{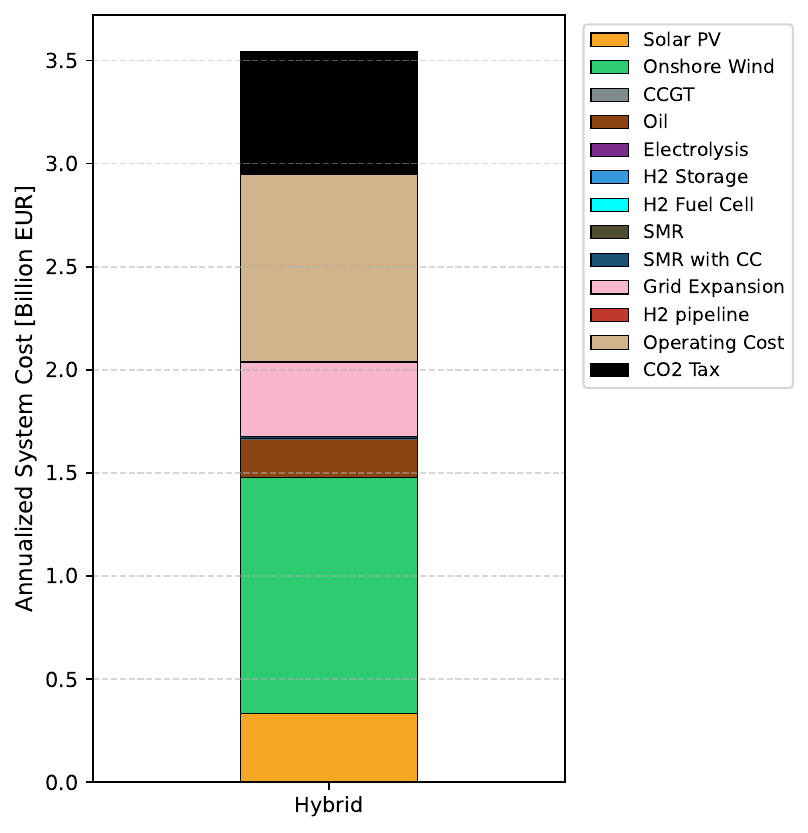}
        \caption{Zero-export baseline case}
        \label{fig:cost_hybrid_no_export}
    \end{subfigure}
\caption{Annualized total system cost breakdown under the hybrid scenario ($100\text{~\texteuro/tonne}$ $\text{CO}_2$ carbon tax): (a) with 70~TWh hydrogen export and (b) zero-export baseline case.}
\label{fig:cost_breakdown_hybrid}
\end{figure}

The annualized system cost structure for the hybrid scenario is shown in Figure~\ref{fig:cost_breakdown_hybrid}, comparing the $70\text{~TWh}$ export case against the zero-export baseline.

\begin{itemize}
    \item {Export Burden and Scale Expansion:} Meeting the 70~TWh external hydrogen demand elevates the total annual system cost to 9.21~Billion~\texteuro/year (9.212~B~\texteuro/yr), representing a $159.93\%$ increase over the zero-export baseline (3.54~Billion~\texteuro/year). This expenditure surge is overwhelmingly driven by natural gas feedstock procurement and operational scaling rather than pure renewable expansion.
    \item {SMR with Carbon Capture Dominance:} To satisfy the export volume under strict carbon penalties without over-investing in dedicated renewable generation, the optimizer relies heavily on Steam Methane Reforming with Carbon Capture (\text{SMR-CC}). Capital allocation for SMR-CC stands as the largest single technology investment at 2.50~Billion~\texteuro/year (2.496~B~\texteuro/yr in export vs.~0.00~B~\texteuro/yr baseline), whereas green electrolysis CAPEX remains negligible (0.0004~Billion~\texteuro/year).
    \item {Operational Cost and Tax Liabilities:} Operating costs constitute the single largest cost component in the export case at 3.60~Billion~\texteuro/year ($39.11\%$ of total system expenditure), reflecting fuel inputs for fossil-based pathways. Meanwhile, the 100~\texteuro/tonne carbon tax liability increases from 0.59~Billion~\texteuro/year in the zero-export baseline to 0.79~Billion~\texteuro/year in the export case, demonstrating that carbon capture deployment effectively caps emission penalties despite the massively scaled hydrogen supply output.
    \item {Renewable Integration Profile:} Onshore wind forms the backbone of electricity CAPEX (1.15~Billion~\texteuro/year under export vs.~1.15~Billion~\texteuro/year baseline), outscaling solar PV investment (0.33~Billion~\texteuro/year across both cases) in both operational regimes.
\end{itemize}

\begin{figure}[htbp]
\centering
\begin{subfigure}[b]{0.40\textwidth}
    \centering
    \includegraphics[width=\textwidth]{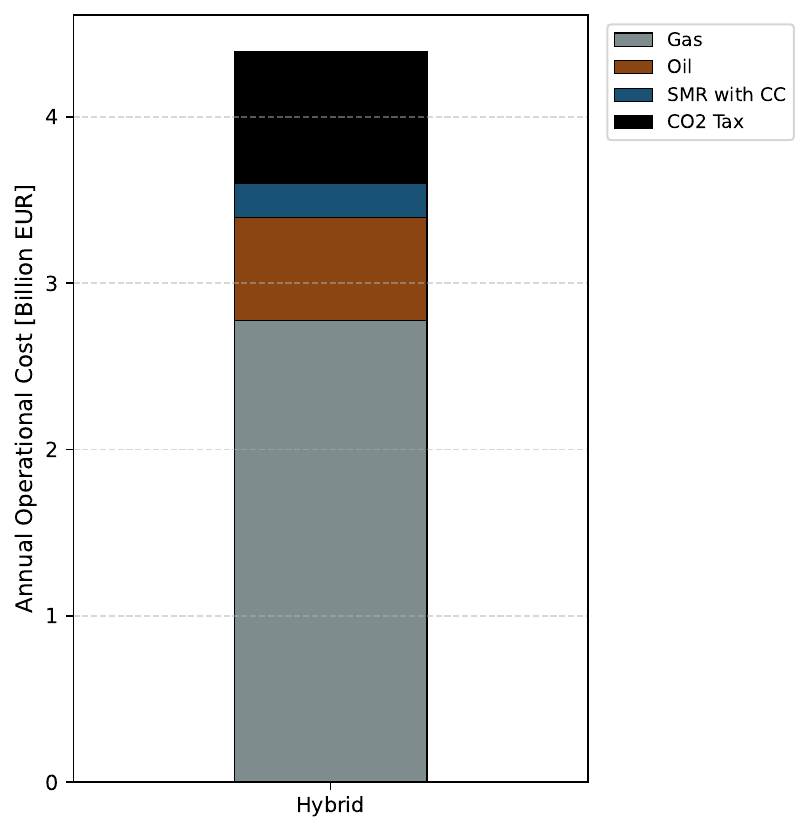}
    \caption{With hydrogen export (70~TWh)}
    \label{fig:hybrid_opex_costs_H2_export}
\end{subfigure}
\hfill
\begin{subfigure}[b]{0.40\textwidth}
    \centering
    \includegraphics[width=\textwidth]{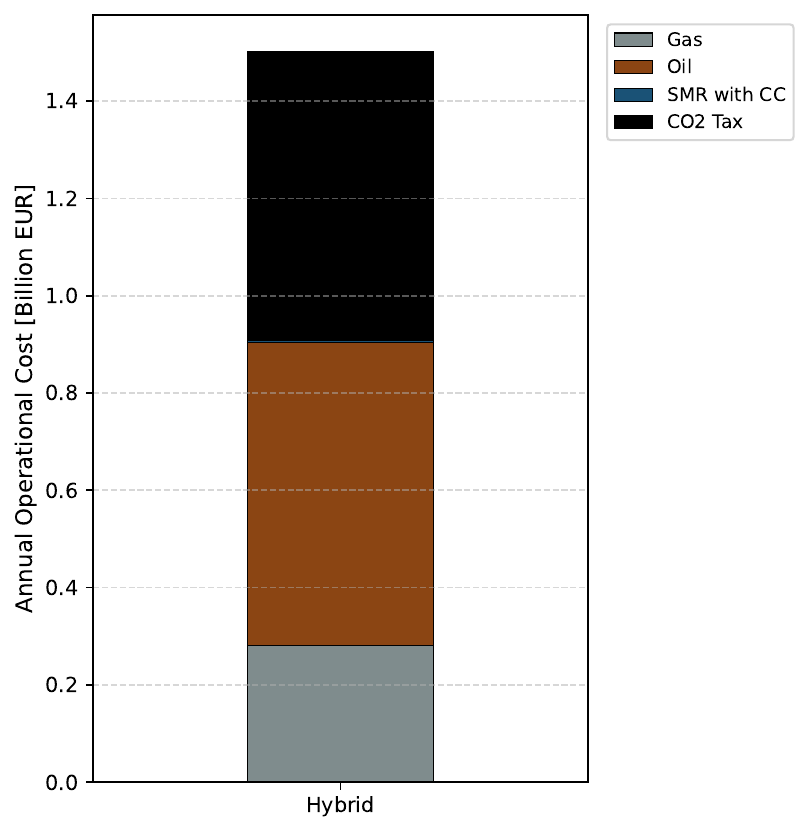}
    \caption{Zero-export baseline case}
    \label{fig:hybrid_opex_costs_base}
\end{subfigure}
\caption{Operational Expenditure (OPEX) breakdown for the hybrid model ($100\text{~\texteuro/tonne}$ $\text{CO}_2$ carbon tax): (a) with 70~TWh hydrogen export and (b) zero-export baseline case.}
\label{fig:hybrid_annual_opex_costs}
\end{figure}

The hybrid OPEX breakdown (Fig.~\ref{fig:hybrid_annual_opex_costs}) illustrates that 70~TWh export demand surges operational expenses from 1.50~Billion~\texteuro/year to 4.40~Billion~\texteuro/year, heavily dominated by natural gas input for SMR with CC (2.78~Billion~\texteuro/year vs.~0.28~Billion~\texteuro/year baseline), alongside consistent oil fuel costs (0.62~Billion~\texteuro/year) and carbon tax liabilities (0.79~Billion~\texteuro/year export vs.~0.59~Billion~\texteuro/year baseline). Notably, primary fuel procurement and carbon penalties overwhelmingly account for the system's operational expenditure, rendering non-fuel OPEX components negligible across both export regimes.

When allowing all hydrogen production technologies to compete directly in a technology-neutral hybrid framework across a range of carbon prices, steam methane reforming with carbon capture (SMR with CC) consistently outcompetes green electrolysis, capturing the dominant share of hydrogen production. This indicates that transitional blue hydrogen offers a lower-cost compliance pathway than pure green hydrogen even under heavy carbon taxation.

\subsection{Levelized cost of hydrogen (LCOH) under hybrid pathway}
To evaluate the economic viability of hydrogen production for export under combined carbon pricing and export constraints, we compute the net levelized cost of hydrogen for the hybrid scenario. This metric provides a standardized basis for evaluating export competitiveness under a $100\text{~\texteuro/tonne}$ $\text{CO}_2$ carbon tax regime.

To isolate the incremental cost of hydrogen production from existing domestic power grid requirements, we utilize a paired-run subtraction method based on Eq.~\ref{Eq:Net-LCOH}. The model is solved under the 70~TWh export demand and compared against a corresponding zero-export baseline.

This subtraction approach ensures that domestic electricity consumers are not modeled as subsidizing the export infrastructure, providing a true marginal production cost for the European market.

\textit{Remark:} The LCOH calculations presented here reflect purely capital and operational production costs and explicitly exclude $\text{CO}_2$ carbon tax liabilities.

\begin{table}[htbp]
\centering
\caption{Summary of $\text{H}_2$ Export costs and net LCOH breakdown for the hybrid pathway}
\label{tab:h2_export_costs_hybrid}
\begin{tabular}{lc}
\toprule
Metric & Hybrid Scenario \\
\midrule
Export Run Cost [Billion \texteuro] & 8.42 \\
Baseline Zero-Export Cost [Billion \texteuro] & 2.95 \\
Net $\text{H}_2$ Cost [Billion \texteuro] & 5.47 \\
Exported $\text{H}_2$ Energy [TWh] & 70.0 \\
\midrule
Net LCOH [\texteuro/MWh] & 78.15 \\
Net LCOH LHV [\texteuro/kg] & 2.60 \\
Net LCOH HHV [\texteuro/kg] & 3.08 \\
\bottomrule
\multicolumn{2}{p{0.65\textwidth}}{\scriptsize \textit{Note:} Net $\text{H}_2$ Cost = Export System Cost -- Baseline Zero-Export Cost; Net LCOH = Net $\text{H}_2$ Cost / Exported $\text{H}_2$ Energy. LCOH figures exclude $\text{CO}_2$ carbon tax liabilities. Conversions assume $\text{H}_2$ Lower Heating Value (LHV) = $0.03333~\text{MWh/kg}$ and Higher Heating Value (HHV) = $0.03941~\text{MWh/kg}$.}
\end{tabular}
\end{table}

As summarized in Table~\ref{tab:h2_export_costs_hybrid}, the hybrid policy environment results in a Net LCOH of $78.15\text{~\texteuro/MWh}$, which translates to $2.60\text{~\texteuro/kg}$ on a Lower Heating Value (LHV) basis and $3.08\text{~\texteuro/kg}$ on a Higher Heating Value (HHV) basis. Subtracting the baseline domestic grid cost ($2.95\text{~Billion~\texteuro}$) from the full export system expenditure ($8.42\text{~Billion~\texteuro}$) yields an incremental export system cost of $5.47\text{~Billion~\texteuro}$ ($5.470\text{~Billion~\texteuro}$).

This unit production cost highlights the economic trade-off of the hybrid regime: while the $100\text{~\texteuro/tonne}$ $\text{CO}_2$ tax restricts unabated fossil options and forces investment in carbon capture infrastructure (\text{SMR-CC}), the ability to share grid flexibility prevents the extreme capital premiums associated with purely off-grid renewable pathways. Consequently, the resulting price point of $2.60\text{~\texteuro/kg LHV}$ positions Libyan hydrogen exports competitively against projected European import cost benchmarks.

\subsection{Net present value (NPV) under hybrid pathway}
Table~\ref{tab:npv_summary_hybrid} presents the financial performance metrics for the hybrid scenario under a dynamic cost-plus pricing framework. A constant revenue margin of $+1.00\text{~\texteuro/kg}$ above the scenario LCOH LHV is applied over the 21-year operational horizon (2030–2050) using a $7.10\%$ real discount rate. While the base LCOH metric reflects unit production costs exclusive of carbon penalties, carbon tax liabilities are separately integrated into the broader system-level cash flow, NPV, IRR, and payback evaluations to ensure comprehensive financial accounting.

\begin{table}[htbp]
\centering
\begin{threeparttable}
\caption{Annual operational cash flows, project return metrics, and multi-year discounted financial evaluation (2030--2050) for the hybrid scenario under a $+1.00\text{~\texteuro/kg}$ revenue margin pricing model and a $100\text{~\texteuro/tonne}$ $\text{CO}_2$ carbon tax ($r = 7.10\%$).}
\label{tab:npv_summary_hybrid}
\small
\setlength{\tabcolsep}{12pt}
\begin{tabular}{lc}
\toprule
Financial Metric & Hybrid Scenario \\ 
\midrule
LCOH LHV Cost [\texteuro/kg] & 2.60 \\
Selling Price (+1 \texteuro/kg) [\texteuro/kg] & 3.60 \\
Export System Cost [B \texteuro/yr] & 8.69 \\
$\text{CO}_2$ Tax Liability [B \texteuro/yr] & 0.79 \\
Annual Total System Cost [B \texteuro/yr] & 9.48 \\
Annual Revenue [B \texteuro/yr] & 7.57 \\
Annual System Net Cash Flow [B \texteuro/yr] & -1.91 \\
Annual Project Net Cash Inflow [B \texteuro/yr] & +2.66 \\
\midrule
Discounted System Cost [B \texteuro] & 101.93 \\
Discounted Gross Revenue [B \texteuro] & 81.38 \\
Total System NPV [B \texteuro] & -20.55 \\
Project IRR (\%) & 12.54\% \\
Simple Payback Period [Years]\tnote{a} & 7.31 \\ 
Discounted Payback Period [Years]\tnote{b} & 10.67 \\ 
\bottomrule
\end{tabular}

\begin{tablenotes}
\footnotesize
\item[*] Revenue margin is added directly to LCOH LHV. Annual total system cost equals Export System Cost plus $\text{CO}_2$ Tax Liability. System annual costs and subsequent cash flows explicitly include the $100\text{~\texteuro/tonne}$ $\text{CO}_2$ carbon tax liability.
\item[a] Simple Payback Period represents the undiscounted capital recovery timeframe in years, calculated as $\text{Simple Payback} = \frac{\text{Net Overnight CAPEX}}{\text{Annual Net Operating Cash Flow}}$.
\item[b] Discounted Payback Period (DPP) accounts for the time value of money ($r = 7.10\%$), evaluated analytically as $\text{DPP} = \frac{-\ln\left(1 - \frac{\text{CAPEX} \cdot r}{\text{Annual Net Cash Flow}}\right)}{\ln(1 + r)}$.
\end{tablenotes}
\end{threeparttable}
\end{table}

Under the combined policy setting, the base LCOH LHV of 2.60~\texteuro/kg establishes a dynamic market selling price of 3.60~\texteuro/kg. At this price level, annual gross revenues reach 7.57~Billion~\texteuro/year. When evaluated against the total annual system expenditure of 9.48~Billion~\texteuro/year (composed of 8.69~Billion~\texteuro/year in export system costs and 0.79~Billion~\texteuro/year in carbon tax liabilities), total system-level net cash flows settle at $-1.91$~Billion~\texteuro/year, yielding a multi-year discounted system NPV of $-20.55$~Billion~\texteuro~(81.38~Billion~\texteuro~discounted revenue vs. 101.93~Billion~\texteuro~discounted system cost).

However, when evaluating the commercial viability of the dedicated hydrogen export infrastructure, the $+1.00$~\texteuro/kg margin over net production costs generates a strong dedicated annual net cash inflow of $+2.66\text{~Billion~\texteuro/year}$. Assessing this project-level cash flow stream against net initial capital outlays yields a Project Internal Rate of Return (IRR) of $12.54\%$ and a payback period of $7.31\text{~years}$.

These results confirm that even under a penalizing $100\text{~\texteuro/tonne}$ carbon tax regime, the hybrid export project remains bankable under a standard cost-plus commercial off-take agreement. The $+1.00\text{~\texteuro/kg}$ premium provides sufficient financial cushion to secure double-digit project returns above the $7.10\%$ hurdle rate.

\subsection{Environmental and resource evaluation under hybrid pathway}
\label{sec:environmental_eval}
This section evaluates the environmental impact and resource demand of the hybrid scenario, focusing on $\text{CO}_2$ emissions balance, carbon capture effectiveness, and fresh water requirements for hydrogen production.

Under the hybrid regime, gross system $\text{CO}_2$ generation is partitioned into net atmospheric emissions (vented) and geologically sequestered emissions (captured via CCS). As detailed in Table~\ref{tab:co2_mass_balance_hybrid}, total gross generation reaches $52.57\text{~Mt}$, of which $44.65\text{~Mt}$ ($84.93\%$) is successfully captured and sequestered. The remaining $7.92\text{~Mt}$ of net atmospheric emissions consists primarily of unabated oil generation ($5.70\text{~Mt}$), uncaptured process emissions from \text{SMR-CC} ($2.03\text{~Mt}$), and residual industrial gas usage ($0.19\text{~Mt}$). This high capture rate highlights how the $100\text{~\texteuro/tonne}$ carbon tax effectively incentivizes deep carbon abatement across the production chain.

\begin{table}[htbp]
\centering
\caption{$\text{CO}_2$ mass balance and vented emissions breakdown for the hybrid scenario.}
\label{tab:co2_mass_balance_hybrid}
\begin{tabular}{lc}
\toprule
$\text{CO}_2$ Accounting Metric / Source & Value \\
\midrule
Net Atmospheric $\text{CO}_2$ Emissions (Vented) & 7.92 Mt \\
Geologically Sequestered $\text{CO}_2$ (Captured) & 44.65 Mt \\
{Total Gross $\text{CO}_2$ Generated (Vented + Stored)} & {52.57 Mt} \\
\midrule
\textit{Breakdown of Vented Emissions by Source:} & \\
~~Oil & 5.70 Mt \\
~~\text{SMR-CC} & 2.03 Mt \\
~~Industrial Gas & 0.19 Mt \\
\bottomrule
\end{tabular}
\end{table}

A critical consideration for hydrogen production in Libya's arid environment is the associated water consumption. Reverse Osmosis (RO) desalination energy overheads and costs are integrated directly into the system model, applying technology-specific water intensities ($10\text{~L/kg}$ for electrolysis, $7.5\text{~L/kg}$ for \text{SMR-CC}, and $5\text{~L/kg}$ for unabated \text{SMR}). 

To support a total hydrogen output of $2.13\text{~Mt}$ ($70.92\text{~TWh}$), total fresh water demand is $15.96\text{~Million~m}^3/\text{year}$ (Table~\ref{tab:water_demand_hybrid}). Despite this volume, desalination imposes a negligible economic and energetic burden on the system: annual desalination costs total $23.94\text{~Million~\texteuro/year}$, accounting for just $0.68\%$ ($0.675\%$) of total system expenditure and adding a marginal cost of only $0.01\text{~\texteuro/kg}_{\text{H}_2}$ ($0.0112\text{~\texteuro/kg}_{\text{H}_2}$). From an energy perspective, the required $55.85\text{~GWh}$ of electricity consumption yields a specific energy overhead of $0.03\text{~kWh/kg}_{\text{H}_2}$ ($0.0262\text{~kWh/kg}_{\text{H}_2}$), representing a mere $0.05\%$ of overall system electricity demand. Consequently, while coastal desalination infrastructure is essential, water supply does not act as an economic or energetic bottleneck for hydrogen export.

\begin{table}[htbp]
\centering
\caption{Dynamic multi-technology water consumption and desalination summary for the hybrid scenario.}
\label{tab:water_demand_hybrid}
\begin{tabular}{lc}
\toprule
Desalination \& Water Consumption Metric & Hybrid Scenario \\
\midrule
Total $\text{H}_2$ Output ($\text{TWh}_{\text{H2}}$ / Million Tonnes) & 70.92 / 2.13 \\
Total Water Demand ($\text{Million~m}^3$) & 15.96 \\
Estimated Desalination Cost ($\text{Million~\texteuro/yr}$) & 23.94 \\
Desalination Cost Share of Total System (\%) & 0.675\% \\
Desalination Marginal Cost ($\text{\texteuro/kg}_{\text{H2}}$) & 0.0112 \\
Desalination Energy Overhead ($\text{GWh}$) & 55.85 \\
Desalination Specific Energy ($\text{kWh/kg}_{\text{H2}}$) & 0.0262 \\
Desalination Share of Total Electricity Demand (\%) & 0.050\% \\
\bottomrule
\multicolumn{2}{p{0.68\textwidth}}{\footnotesize \textit{Note:} Based on an assumed seawater desalination cost of $1.50\text{~\texteuro/m}^3$ and technology-specific water intensities.}
\end{tabular}
\end{table}

\section{Conclusion}
\label{sec:conclusion}

This study introduces \textsf{PyPSA-Libya}, the first open-source, high-resolution energy system model designed to evaluate Libya’s national decarbonization pathways and regional hydrogen export potential to Europe. 

Our techno-economic analysis demonstrates that Libya's world-class solar and wind resources enable highly competitive green hydrogen production under a dedicated 100\% renewable electrolysis strategy. 

However, when allowing all hydrogen production technologies to compete directly within a technology-neutral hybrid framework across varied carbon prices, steam methane reforming with carbon capture (SMR with CC) consistently outcompetes green electrolysis, capturing the dominant share of production. This indicates that transitional blue hydrogen offers a lower-cost compliance pathway than pure green hydrogen, even under progressive carbon taxation. 

Unlike European production hubs subject to high international natural gas price volatility, Libya's status as a major domestic hydrocarbon producer insulates local SMR-with-CC pathways from external feedstock shocks. This inherent resource insulation further strengthens the economic attractiveness and long-term viability of the blue hydrogen export framework.

While unconstrained grey hydrogen strategies face severe long-term financial liabilities under European carbon pricing regimes, the economic viability of green electrolysis remains closely tied to capital cost trajectories and temporal matching constraints. 

Furthermore, integrating sector-coupling mechanisms, such as hydrogen systems, provides critical operational flexibility to handle high shares of variable renewable generation and residual system variability at the lowest total system cost—effectively displacing the need for capital-intensive short-duration battery storage.

As with energy planning frameworks generally, default macro-energy optimization architectures present inherent limitations by evaluating technology placement strictly through a techno-economic lens, often omitting local physical constraints such as inland water availability for electrolysis. This highlights the vital importance of open-source modeling frameworks, which empower researchers to modify technology allocation logic and enforce geographic realism. 

Consequently, \textsf{PyPSA-Libya} serves as a decision-support tool whose outputs are conditional on data resolution and techno-economic assumptions; findings should be interpreted alongside detailed engineering and stakeholder assessments.

Ultimately, while Libya possesses a strong strategic advantage to supply decarbonized fuels to the European market, realizing this potential will strictly depend on early cross-border coordination, resolving domestic grid transmission bottlenecks, and securing targeted cross-Mediterranean infrastructure investments.

\section*{Code and data availability}
The model code is available at \url{https://github.com/MohamedAbuella/PyPSA-Libya}.

\appendix
\section{Optimization problem scale and computational complexity}
\label{sec:appendix_problem_scale}

\setcounter{table}{0}
\renewcommand{\thetable}{A.\arabic{table}}
\setcounter{equation}{0}
\renewcommand{\theequation}{A.\arabic{equation}}
\setcounter{figure}{0}
\renewcommand{\thefigure}{A.\arabic{figure}}

The sector-coupled optimization model for Libya is formulated as a Linear Program (LP) using \textsf{linopy} with 2,920 time snapshots (8,760 hours at a 3-hour resolution) across 12 spatial clusters. Table~\ref{tab:problem_dimensions} summarizes the structural dimensions of the linear programming matrices for the three analyzed scenarios.

\begin{table}[htbp]
\centering
\caption{Linear programming matrix dimensions across the modeled scenarios.}
\label{tab:problem_dimensions}
\begin{tabular}{lrrr}
\toprule
{Scenario} & {Decision Variables} & {Constraints} & {Non-Zero Entries} \\ \midrule
Scenario A & 2,228,414 & 4,596,589 & 9,611,971 \\
Scenario B & 2,263,466 & 4,666,680 & 9,752,143 \\
Scenario C & 2,298,518 & 4,736,772 & 9,962,395 \\ \bottomrule
\end{tabular}
\end{table}

The primary driver of the dimensional variation across scenarios is the number of enabled hydrogen production technologies. In PyPSA, each active technology instantiated across network nodes generates both capacity investment variables and operational dispatch decision variables for every time step. Scenario A restricts production exclusively to \textsf{H2 Electrolysis}, Scenario B includes \textsf{H2 Electrolysis} and \textsf{SMR CC}, and Scenario C permits all three options (\textsf{H2 Electrolysis}, \textsf{SMR}, and \textsf{SMR CC}). Excluding a technology skips component instantiation; each additional SMR technology adds exactly 35,052 decision variables ($2,920 \text{snapshots} \times 12 \text{nodes} + \text{investment variables}$) and $70,092$ constraints to the matrix.


\bibliographystyle{elsarticle-num}
\bibliography{Libya_H2}

\end{document}